\documentclass[prb,twocolumn,superscriptaddress]{revtex4-2}
\usepackage{amsmath}
\usepackage{color}
\usepackage{bbm}
\usepackage{amssymb}
\usepackage{epsfig}
\usepackage{multirow}
\usepackage{amsbsy}
\usepackage{array}
\usepackage{diagbox}
\usepackage{bm}
\usepackage{tikz}
\usepackage{extarrows}
\usepackage{graphicx}
\usepackage{appendix}
\usepackage{txfonts}
\usepackage[normalem]{ulem}
\usepackage[colorlinks=true,linkcolor=blue,citecolor=blue,urlcolor=blue,bookmarks=false]{hyperref}

\newcommand{\dd}[0]{\mathrm{d}}
\newcommand{\ii}[0]{\mathrm{i}}
\newcommand{\ee}[0]{\mathrm{e}}

\begin{document}

\title{Apparent double-$T_c$ from a single BKT transition in anisotropic phase-only models}

\author{Pei-Yuan Cai}
\affiliation{Institute of Physics, Chinese Academy of Sciences, Beijing 100190, China}
\affiliation{School of Physical Sciences, University of Chinese Academy of Sciences, Beijing 100190, China}

\author{Yi Zhou}
\email{yizhou@iphy.ac.cn}
\affiliation{Institute of Physics, Chinese Academy of Sciences, Beijing 100190, China}

\date{\today}

\begin{abstract}
Transport experiments on two-dimensional superconductors often yield direction-dependent transition temperatures, raising the question of whether such a ``double-$T_c$'' reflects a true thermodynamic splitting or a transport artifact. To establish a baseline, we study a minimal anisotropic phase-only Josephson-junction array in equilibrium and under resistively shunted junction dynamics with fluctuating twist boundary conditions. The equilibrium model exhibits a single Berezinskii--Kosterlitz--Thouless (BKT) transition. Out of equilibrium, anisotropic Josephson couplings and anisotropic dissipation reshape the linear $R$--$T$ curves in a finite-size, finite-current crossover regime, so that curve-shape criteria such as Halperin--Nelson fits and fixed-resistance thresholds yield an apparent double-$T_c$. In contrast, critical-scaling criteria---the universal exponent $\alpha=3$ and dynamic finite-size scaling---remain consistent with the single $T_{\mathrm{BKT}}$. A robust splitting that persists in the nonlinear critical scaling, such as that recently reported at KTaO$_3$ interfaces, therefore points to physics beyond this clean anisotropic baseline.
\end{abstract}

\maketitle

\section{Introduction}\label{sec:introduction}
 In two-dimensional (2D) superconductors, phase coherence is governed not merely by the mean-field onset of the order-parameter amplitude, but predominantly by phase fluctuations and topological defects. In the standard Berezinskii--Kosterlitz--Thouless (BKT) paradigm, the true thermodynamic transition is driven by the unbinding of integer vortex--antivortex pairs, marked by a universal jump in the renormalized superfluid stiffness (or helicity modulus) \cite{berezinskii1971destruction,kosterlitz1973ordering,kosterlitz1974critical,nelson1977universal,fisher1973helicity,minnhagen1987two}. Experimentally, however, the thermodynamic transition must be inferred from nonequilibrium transport measurements, typically manifesting as nonlinear current--voltage ($I$--$V$) characteristics and a rapid drop in linear resistance--temperature ($R$--$T$) curves \cite{ambegaokar1978dissipation,ambegaokar1980dynamics,halperin1979resistive}. Because finite sample sizes and probe currents inherently cut off critical divergences, transport-extracted temperatures obtained from these operational metrics are not automatically identical to the underlying equilibrium BKT temperature $T_{\text{BKT}}$. 
 
 This distinction between thermodynamics and transport becomes paramount when the macroscopic response is strongly direction-dependent. Recent transport experiments on oxide heterointerfaces, specifically EuO/KTaO$_3$(110) and (111), have brought this issue into sharp focus \cite{hua2024superconducting, Huang2026}. In these systems, the superconducting transition temperature and upper critical fields depend distinctly on the in-plane current direction. Interpretations of such an apparent ``double-$T_c$'' often invoke exotic, spatially segregated physics, such as quasi-one-dimensional superconducting stripes where phase coherence develops earlier along one crystallographic axis than another. The fundamental question is therefore: does a direction-dependent transport-extracted temperature necessarily imply a split thermodynamic transition (or exotic spatially decoupled phases), or can it emerge as a transport artifact in a system with a single, uniform BKT transition? 
 
 Recently, several theoretical frameworks have been proposed to explain this striking experimental phenomenon. For instance, Li, Kivelson, and Lee demonstrated that an infinitely anisotropic superconducting phase can emerge via the unbinding of half-vortices \cite{li2024theory}. Concurrently, Xu, Jiang, and Hu highlighted the critical role of anisotropic vortex motion induced by asymmetric pinning landscapes \cite{xu2025anisotropic}. These sophisticated models provide compelling mechanisms for true directional decoupling. However, they also motivate a foundational theoretical question: before invoking higher-order topological defects or explicit disorder potentials, what is the baseline expectation? Specifically, how much of the apparent double-$T_c$ phenomenology naturally arises from the simplest clean, anisotropic phase-only model? 
 
 In this work, we address this by studying a coarse-grained anisotropic Josephson-junction array (JJA) in both equilibrium and nonequilibrium settings. In equilibrium, the model reduces to an anisotropic XY model, and the true thermodynamic transition is determined from the helicity modulus. Out of equilibrium, we evolve the same phase-only Hamiltonian using resistively shunted junction dynamics (RSJD) under fluctuating twist boundary conditions (FTBC) \cite{mon1989phase,chung1989dynamical,kim1999vortex,choi2000boundary,medvedyeva2000analysis}. This unified approach allows us to separate the equilibrium BKT temperature from the criterion-dependent transport-extracted temperatures obtained from finite-current $E$--$j$ and $r_{\mathrm{lin}}$--$T$ curves, as the numerical counterparts of experimental $I$--$V$ and $R$--$T$ transport measurements. 
 
 Our results demonstrate that a purely continuous anisotropic model supports only a single thermodynamic BKT transition, dictated by the geometric mean of the directional stiffnesses. However, finite-current $r_{\mathrm{lin}}$--$T$ curves can readily display an apparent, direction-dependent splitting of transport-extracted temperatures. We show that this splitting is not a thermodynamic reality, but a crossover effect. Because finite system sizes and probe currents truncate the critical divergence of the correlation length, $R$--$T$ curve fits (such as the Halperin--Nelson formula) are forced into a higher-temperature crossover regime. In this regime, non-universal anisotropic Josephson couplings and anisotropic dissipation---which phenomenologically captures anisotropic viscous vortex drag---reshape the resistive curves differently along orthogonal axes, leading curve-shape criteria to yield two distinct apparent transport-extracted temperatures. 
 
 Crucially, we find that criteria tied directly to critical transport scaling are much less sensitive to this crossover artifact. The nonlinear $I$--$V$ exponent criterion ($\alpha=3$) and dynamic finite-size scaling (FSS) are less affected by the curve-reshaping and remain consistent with the single equilibrium BKT temperature $T_{\text{BKT}}$. By contrasting the failure of $R$--$T$ fits with the robustness of $I$--$V$ scaling, our model provides a useful diagnostic comparison for experimentalists. In the scenario captured by the present clean baseline, an apparent directional splitting in the transport-extracted temperatures from $R$--$T$ curves is not accompanied by a resolvable splitting in the critical scaling criteria. Conversely, if the splitting persists in the $\alpha=3$ critical scaling---as is notably the case in recent EuO/KTaO$_3$(111) measurements \cite{Huang2026}---it suggests that the system harbors exotic physics beyond the standard continuous anisotropic BKT paradigm. 
 
 The rest of the paper is organized as follows. Section~\ref{sec:model} introduces the anisotropic JJA model and the observables used in the equilibrium and nonequilibrium analyses. Section~\ref{sec:results} presents the numerical results, contrasting the single thermodynamic transition with the bifurcated transport extractions. Section~\ref{sec:discussion} summarizes the physical picture, discusses the mechanism of anisotropic curve reshaping, and outlines the scope and application of this theoretical baseline.

\section{Model and numerical observables}\label{sec:model}
	
	\subsection{Effective anisotropic JJA model}
 \begin{figure}[tb]
		\centering
		\includegraphics[width=1.\linewidth]{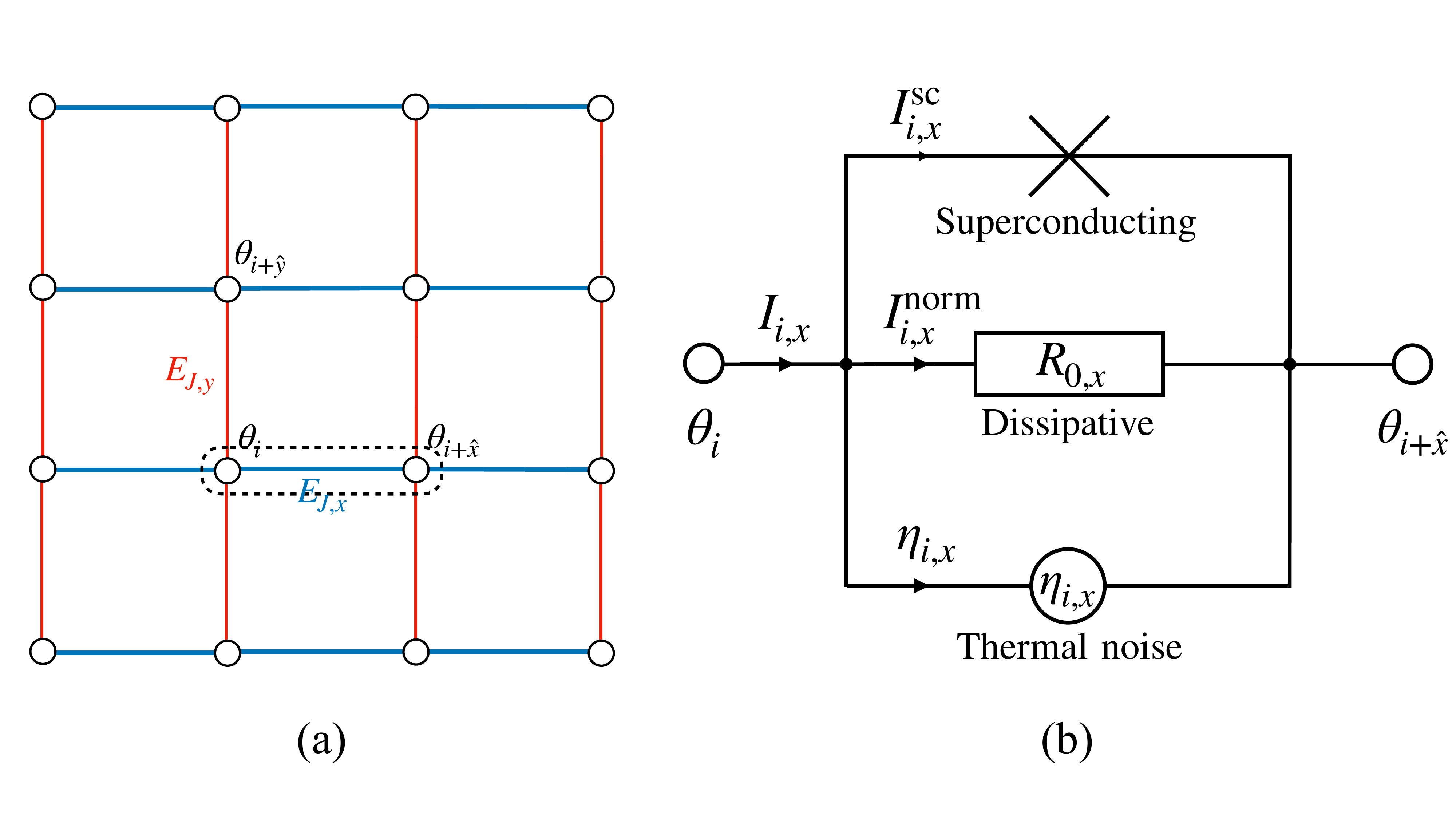}
		\caption{Schematic of the anisotropic phase-only model and the RSJ dynamics. (a) Coarse-grained anisotropic JJA/XY model on a square lattice. The superconducting phase on site $i$ is denoted by $\theta_i$. Horizontal and vertical bonds represent the effective Josephson stiffnesses $E_{J,x}=qE_*$ and $E_{J,y}=(1-q)E_*$, respectively. (b) RSJ representation example of the boxed $x$-directed bond in (a). The total bond current $I_{i,x}$ is decomposed into three channels given in Eq.~\eqref{eq:bond_current}. The corresponding $y$-bond is obtained by replacing $x$ with $y$. }
 \label{fig:model}
	\end{figure}
 
	We consider a two-dimensional superconducting order parameter
	\[
	\Psi(\mathbf r)=|\Psi(\mathbf r)|\ee^{\ii\theta(\mathbf r)},
	\qquad \mathbf r=(x,y).
	\]
	In the temperature window of interest, namely below the mean-field pairing scale $T_{\mathrm{MF}}$ and near the BKT transition, amplitude fluctuations are assumed to be less important than phase fluctuations. The long-wavelength free energy is then approximated by the phase-only form
	\[
	F[\theta]=\frac12\int \dd^2\mathbf r\sum_{\mu=x,y}\rho_\mu(\partial_\mu\theta)^2,
	\]
	where $\rho_\mu$ is the phase stiffness along direction $\mu$.
	
	After coarse graining onto an $L\times L$ lattice of superconducting islands [see Fig.~\ref{fig:model}.(a)], we denote the phase on site $i$ by $\theta_i$ and write the effective Hamiltonian as
	\begin{equation}\label{eq:H_JJA}
		H=-\sum_i\sum_{\mu=x,y}E_{J,\mu}\cos\phi_{i,\mu},
	\end{equation}
	with the gauge-invariant phase difference
	\begin{equation}\label{eq:gaugeinvphi}
		\phi_{i,\mu}=\theta_{i+\hat\mu}-\theta_i+A_{i,\mu}.
	\end{equation}
	For a literal Josephson-junction array, $E_{J,\mu}=\hbar I_{c,\mu}/(2e)$ is the Josephson coupling energy of a junction. In the present coarse-grained description of a continuous two-dimensional superconductor, $E_{J,\mu}$ should be viewed more generally as an effective bond stiffness. It is the local energy cost of twisting the superconducting phase across a coarse-grained bond in direction $\mu$. 
	
	For the zero-field problem studied here, the link twist $A_{i,\mu}$ is introduced only to encode a boundary twist. A total twist $\Theta_\mu$ along direction $\mu$ may be represented either by a uniform link twist $A_{i,\mu}=A_\mu=\Theta_\mu/L$ with periodic boundary conditions (PBC) on $\theta_i$, or equivalently by gauging away $A_\mu$ and imposing twisted boundary conditions (TBC) on the phase variables. The two descriptions are gauge equivalent and lead to the same equilibrium thermodynamics.
	
	To parameterize the anisotropy, throughout this work we use: 
	\begin{equation}\label{eq:q_parameterization}
		E_{J,x}=qE_*, \qquad E_{J,y}=(1-q)E_*, \qquad \frac{1}{2} \leq q < 1,
	\end{equation}
	where $E_*=E_{J,x}+E_{J,y}$ sets the overall energy scale. With $k_B=1$, temperature is measured in the same units. We exclude $q=1$ because the system turns into one-dimensional chains there and no finite-temperature BKT transition remains. The isotropic limit corresponds to $q=1/2$, for which Eq.~(\ref{eq:H_JJA}) becomes
	\[
	H=-\frac{E_*}{2}\sum_i\bigl(\cos\phi_{i,x}+\cos\phi_{i,y}\bigr).
	\]
	Comparison with the standard isotropic XY Hamiltonian $H=-J\sum_i(\cos\phi_{i,x}+\cos\phi_{i,y})$ shows that the isotropic coupling is $J=E_*/2$. Thus, if one sets $E_*=1$, the isotropic transition temperature is $T_{\mathrm{BKT}}\approx 0.446$, i.e., one half of the usual unit-coupling value $\approx 0.893$\cite{olsson1995monte1,olsson1995monte2,weber1988monte,hubscher2013stiffness}. 
	
	The model in Eq.~(\ref{eq:H_JJA}) provides the common starting point for our analysis. In the classical equilibrium limit, it is equivalent to the two-dimensional anisotropic XY model\cite{beasley1979possibility,ohta1979xy,teitel1983phase}. Out of equilibrium, the nonlinear transport can be studied by the RSJD, which governs the phase evolution under external driving currents and thermal fluctuations \cite{mon1989phase,chung1989dynamical,kim1999vortex,choi2000boundary,medvedyeva2000analysis}. 
	
	\subsection{Equilibrium: universal jump and helicity modulus}\label{subsection:equilibrium}
	
	\subsubsection{Single BKT transition from the long-wavelength theory}
	The equilibrium question is whether anisotropy can split the BKT transition into two distinct transitions. To address this, we expand Eq.~(\ref{eq:H_JJA}) for smoothly varying phase configurations with $A_{i,\mu}=0$. The spin-wave part of the dimensionless energy is
	\begin{equation}\label{eq:free_energy_spin_wave}
		\frac{F_{\mathrm{sw}}}{T}=
		\frac12\int \dd^2\mathbf r
		\left[K_x(\partial_x\theta)^2+K_y(\partial_y\theta)^2\right],
	\end{equation}
	where
	\begin{equation}
		K_x=\frac{E_{J,x}}{T},
		\qquad
		K_y=\frac{E_{J,y}}{T}
	\end{equation}
	are the bare dimensionless stiffnesses.
	
	The anisotropy can be removed by the coordinate rescaling
	\begin{equation}
		x'=\left(\frac{K_y}{K_x}\right)^{1/4}x,
		\qquad
		y'=\left(\frac{K_x}{K_y}\right)^{1/4}y,
	\end{equation}
	which transforms Eq.~(\ref{eq:free_energy_spin_wave}) into
	\begin{equation}\label{eq:isotropized_spin_wave}
		\frac{F_{\mathrm{sw}}}{T}=
		\frac{\bar K_0}{2}\int \dd^2\mathbf r'\,(\nabla'\theta)^2,
	\end{equation}
	with the effective bare stiffness
	\begin{equation}
		\bar K_0=\sqrt{K_xK_y}=\frac{\sqrt{E_{J,x}E_{J,y}}}{T}.
	\end{equation}
	The long-wavelength theory therefore depends on the anisotropy only through the geometric mean $\sqrt{K_xK_y}$. The standard BKT renormalization-group flow\cite{kosterlitz1973ordering,kosterlitz1974critical,nelson1977universal,jose1977renormalization} then applies with $K\to\bar K$:
	\begin{equation}\label{eq:RGflowK}
		\frac{\dd \bar K^{-1}}{\dd l}=4\pi^3y^2+\mathcal O(y^4),
	\end{equation}
	\begin{equation}\label{eq:RGflowy}
		\frac{\dd y}{\dd l}=\bigl(2-\pi\bar K\bigr)y+\mathcal O(y^3),
	\end{equation}
	where $y$ is the vortex fugacity. Along the critical trajectory, the renormalized stiffness approaches the fixed point\cite{nelson1977universal}
	\begin{equation}
		\bar K_R(T_{\mathrm{BKT}}^-)=\frac{2}{\pi}.
	\end{equation}
	Equivalently, in terms of the renormalized effective helicity modulus
	\begin{equation}
		\bar\Upsilon_R\equiv T\bar K_R,
	\end{equation}
	one obtains the Nelson--Kosterlitz jump condition at the BKT transition temperature\cite{ohta1979xy,teitel1983phase,weber1988monte}
	\begin{equation}\label{eq:Helicity_jump}
		\bar\Upsilon_R(T_{\mathrm{BKT}}^-)=\frac{2T_{\mathrm{BKT}}}{\pi}.
	\end{equation}
	Thus the anisotropy changes the stiffness scale, but the BKT transition is still controlled by a single critical temperature. 
	
	\subsubsection{Directional helicity modulus}
	In the numerical simulations we evaluate the directional helicity moduli, which are defined by
	\begin{equation}\label{eq:Fisher_formula}
		\Upsilon_\mu(L,T) \equiv \left.\frac{\partial^2 F}{\partial {\Phi_\mu}^2}\right|_{{\Phi_\mu}=0},
	\end{equation}
	as the free-energy response to an infinitesimal extra boundary twist $\Phi_\mu$ along the $\mu$ direction\cite{fisher1973helicity,ohta1979xy,teitel1983phase} under given system size $L$, temperature $T$ and most importantly, under PBC, i.e., $\Theta_\mu=0$. It is convenient to introduce the twist per bond
	\begin{equation}
		\delta_\mu\equiv{\Phi_\mu}/L,
	\end{equation}
	so that
	\begin{equation}
		\Upsilon_\mu(L,T) = \frac{1}{L^2} \left.\frac{\partial^2F}{\partial\delta_\mu^2}\right|_{\delta_\mu=0}.
	\end{equation}
	For the Hamiltonian in Eq.~(\ref{eq:H_JJA}), this gives
	\begin{equation}\label{eq:helicity_modulus_MC}
		\Upsilon_\mu(L,T)=\frac{1}{L^2}\langle C_\mu\rangle-\frac{\beta}{L^2}\left[\langle S_\mu^2\rangle-\langle S_\mu\rangle^2\right],
	\end{equation}
	where
	\[
	S_\mu\equiv\sum_iE_{J,\mu}\sin\phi_{i,\mu},
	\qquad
	C_\mu\equiv\sum_iE_{J,\mu}\cos\phi_{i,\mu}.
	\]
	Here the gauge-invariant phase difference becomes $\phi_{i,\mu}=\theta_{i+\hat{\mu}}-\theta_i$ under $A_\mu=\Theta_\mu/L=0$. The detailed derivation of Eq.~\eqref{eq:helicity_modulus_MC} is given in Appendix~\ref{app:helicity_modulus}.
	
	The quantity entering the equilibrium BKT criterion is the geometric mean
	\begin{equation}
		\bar\Upsilon(L,T)\equiv\sqrt{\Upsilon_x(L,T)\Upsilon_y(L,T)}.
	\end{equation}
	For each system size $L$, we define $T^*(L)$ by the crossing
	\begin{equation}
		\bar\Upsilon(L,T)=\frac{2T}{\pi},
	\end{equation}
	and estimate the equilibrium BKT temperature from the finite-size form
	\begin{equation}\label{eq:BKT_finite_size}
		T^*(L)=T_{\mathrm{BKT}}+\frac{A}{\ln^2(bL)},
	\end{equation}
	with $A$ and $b$ fitting parameters \cite{weber1988monte,schultka1994finite,olsson1995monte1,olsson1995monte2}. In the present work, $T_{\mathrm{BKT}}$ is referred to as the equilibrium BKT temperature, namely our finite-size estimate of the thermodynamic BKT transition temperature. 

 We emphasize that the helicity modulus (Eq.~\eqref{eq:Fisher_formula}) is defined under zero boundary twist. Accordingly, the BKT criterion in Eq.~\eqref{eq:Helicity_jump} and the finite-size form in Eq.~\eqref{eq:BKT_finite_size} are applied only to the system under zero boundary twist. At a finite boundary twist, one may instead define a finite-twist stiffness response at $\Theta_\mu\neq0$. This quantity is not identical to Eq.~\eqref{eq:Fisher_formula} and is not used to determine $T_{\mathrm{BKT}}$. Throughout the equilibrium analysis the boundary twist is fixed at zero, and finite-twist consistency checks (which require a modified cluster update) are presented in Appendix~\ref{app:helicity_modulus_finite_twist}. 
	
	\subsection{Nonequilibrium dynamics: RSJD under FTBC}
	\subsubsection{Equations of motion}
	To study finite-current transport, we evolve the same phase-only model with RSJD under FTBC \cite{lobb1983theoretical,mon1989phase,chung1989dynamical,kim1999vortex,choi2000boundary,medvedyeva2000analysis}. In FTBC the local phases remain periodic,
	\[
	\theta_{i+L\hat x}=\theta_i, \qquad \theta_{i+L\hat y}=\theta_i,
	\]
	while the uniform twist variables $\Delta_\mu$ are absorbed into the bond phase differences,
	\begin{equation}\label{eq:gaugeinvphi_ftbc}
		\phi_{i,\mu}=\theta_{i+\hat\mu}-\theta_i-\Delta_\mu.
	\end{equation}
	The ``fluctuating twist'' means that $\Delta_\mu$ will also be evolved in every time step when solving the Langevin equations. 
	
 The resistively shunted junction interpretation of a single bond is shown in Fig.~\ref{fig:model}(b). On each bond, the total current is written as\cite{stewart1968current,mccumber1968effect}
	\begin{equation}\label{eq:bond_current}
		I_{i,\mu}=I_{i,\mu}^{\mathrm{sc}}+I_{i,\mu}^{\mathrm{norm}}+\eta_{i,\mu}.
	\end{equation}
	The three terms are interpreted as follows.
	\begin{enumerate}
		\item \textit{Superconducting channel.} Following the current-phase relation of a Josephson junction\cite{golubov2004current}, the superconducting current is given by 
		\begin{equation}
			I_{i,\mu}^{\mathrm{sc}}
			=\frac{2e}{\hbar}\frac{\partial}{\partial\phi_{i,\mu}}
			\bigl(-E_{J,\mu}\cos\phi_{i,\mu}\bigr)
			=I_{c,\mu}\sin\phi_{i,\mu},
		\end{equation}
		with $I_{c,\mu}=(2e/\hbar)E_{J,\mu}$.
		
		\item \textit{Dissipative channel.} The voltage across a bond is related to the phase by the Josephson relation, $V_{i,\mu}=(\hbar/2e)\dot\phi_{i,\mu}$. The dissipative current is modeled as an Ohmic shunt, 
		\begin{equation}
			I_{i,\mu}^{\mathrm{norm}}=\frac{V_{i,\mu}}{R_{0,\mu}}=\frac{\hbar}{2eR_{0,\mu}}\dot\phi_{i,\mu},
		\end{equation}
		where $R_{0,\mu}$ is the coarse-grained bond shunt resistance along direction $\mu$. 
		
		\item \textit{Thermal noise channel.} Following the Langevin formalism introduced by Ambegaokar and Halperin \cite{ambegaokar1969voltage}, the noise current satisfies the fluctuation--dissipation theorem, 
		\begin{equation}
			\langle\eta_{i,\mu}(t)\rangle=0,
		\end{equation}
		\begin{equation}
			\langle\eta_{i,\mu}(t)\eta_{j,\nu}(t')\rangle=
			\frac{2k_BT^{\mathrm{phys}}}{R_{0,\mu}}\delta_{ij}\delta_{\mu\nu}\delta(t-t').
		\end{equation}
	\end{enumerate}
	
	The equations of motion follow from Kirchhoff current conservation at every site together with the condition that the spatially averaged current equals the applied current density\cite{stewart1968current,mccumber1968effect,ambegaokar1969voltage,lobb1983theoretical,chung1989dynamical}. Their derivation is summarized in Appendix~\ref{app:RSJD_EoM}. After introducing the dimensionless variables defined there, we can obtain the equations of motion for the phase and twist variables: 
	\begin{equation}\label{eq:RSJD_EoM_theta_dimensionless}
		\dot\theta_i=
		\sum_jG_{ij}^{(r_0)}
		\sum_{\mu=x,y}
		\left[
		\mathcal J_{j,\mu}^{\mathrm{sc}}-\mathcal J_{j-\hat\mu,\mu}^{\mathrm{sc}}
		+\zeta_{j,\mu}-\zeta_{j-\hat\mu,\mu}
		\right],
	\end{equation}
	\begin{equation}\label{eq:RSJD_EoM_delta_dimensionless}
		\dot\Delta_\mu=
		r_{0,\mu}
		\left[
		\frac{1}{L^2}\sum_i\mathcal J_{i,\mu}^{\mathrm{sc}}
		+\zeta_{\Delta_\mu}-j_\mu
		\right],
		\qquad \mu=x,y,
	\end{equation}
	where the dot denotes differentiation with respect to dimensionless time $\tau$, and $G_{ij}^{(r_0)}$ is the lattice Green's function of the weighted discrete Laplacian. The dimensionless superconducting current is given by
	\begin{equation}
		\mathcal J_{i,\mu}^{\mathrm{sc}}=J_\mu\sin\phi_{i,\mu},
		\qquad
		J_\mu\equiv\frac{E_{J,\mu}}{E_*},
	\end{equation}
	and the dimensionless shunt resistance $r_{0,\mu}$ and spatially averaged noise current $\zeta_{\Delta_\mu}$ are: 
	\begin{equation}
		r_{0,\mu}\equiv\frac{R_{0,\mu}}{R_*},
		\qquad
		\zeta_{\Delta_\mu}\equiv\frac{1}{L^2}\sum_i\zeta_{i,\mu}.
	\end{equation}
	The dimensionless noise satisfies
	\begin{equation}
		\langle\zeta_{i,\mu}(\tau)\zeta_{j,\nu}(\tau')\rangle=
		\frac{2T}{r_{0,\mu}}\delta_{ij}\delta_{\mu\nu}\delta(\tau-\tau'),
	\end{equation}
	\begin{equation}
		\langle\zeta_{\Delta_\mu}(\tau)\zeta_{\Delta_\nu}(\tau')\rangle=
		\frac{2T}{r_{0,\mu}L^2}\delta_{\mu\nu}\delta(\tau-\tau').
	\end{equation}
	
	The macroscopic voltage drop along direction $\mu$ follows from the Josephson relation, and the corresponding electric field is given by: 
	\begin{equation}
		V_\mu=-\frac{\hbar L}{2e}\dot\Delta_\mu,
		\qquad
		E_\mu^{\mathrm{phys}}=-\frac{\hbar}{2e}\dot\Delta_\mu,
	\end{equation}
	so that the dimensionless electric field measured in the simulation is the long-time average of $\dot{\Delta}_{\mu}$: 
	\begin{equation}\label{eq:electric_field_averaged}
		E_\mu=-\langle\dot\Delta_\mu\rangle_\tau.
	\end{equation}
	These equations provide the basis for all transport calculations below.

	Throughout the paper, we use $E$--$j$ and $r_{\mathrm{lin}}$--$T$ for the dimensionless quantities computed in the simulations. We also refer to them as the numerical counterparts of current--voltage ($I$--$V$) and resistance--temperature ($R$--$T$) measurements, respectively. For a fixed square geometry, $V_\mu$ and $I_\mu$ differ from $E_\mu$ and $j_\mu$ only by geometric factors. Therefore the nonlinear exponent, scaling collapse, and extracted temperature scales are unaffected by this change of notation. 

\section{Numerical simulations, results, and analysis}\label{sec:results}
Before presenting the numerical results, we clarify the terminology used for temperature scales. The temperature obtained from the equilibrium helicity-modulus analysis will be called the \emph{equilibrium BKT temperature}. It is our finite-size estimate of the thermodynamic BKT transition temperature of the underlying anisotropic XY model. By contrast, all temperatures obtained from nonequilibrium RSJD transport data will be called \emph{transport-extracted temperatures}. These are criterion-dependent operational quantities, analogous to transition temperatures inferred from experimental $I$--$V$ or $R$--$T$ measurements, and need not coincide with the equilibrium BKT temperature in a finite-size, finite-current setting. Within the transport-extracted temperatures, we further distinguish critical-scaling criteria, such as the $\alpha=3$ criterion and dynamic finite-size scaling, from curve-shape criteria, such as Halperin--Nelson fits and fixed-resistance thresholds.

\subsection{Equilibrium}
For equilibrium calculations we simulate Eq.~(\ref{eq:H_JJA}) on $L\times L$ lattices with $L=60$, $80$, $100$, $160$, and $200$ using Wolff cluster updates \cite{wolff1989collective}. We study $q=0.5$, $0.6$, $0.7$, $0.8$, and $0.9$.

\begin{figure}[tb]
 \centering
 \includegraphics[width=1.\linewidth]{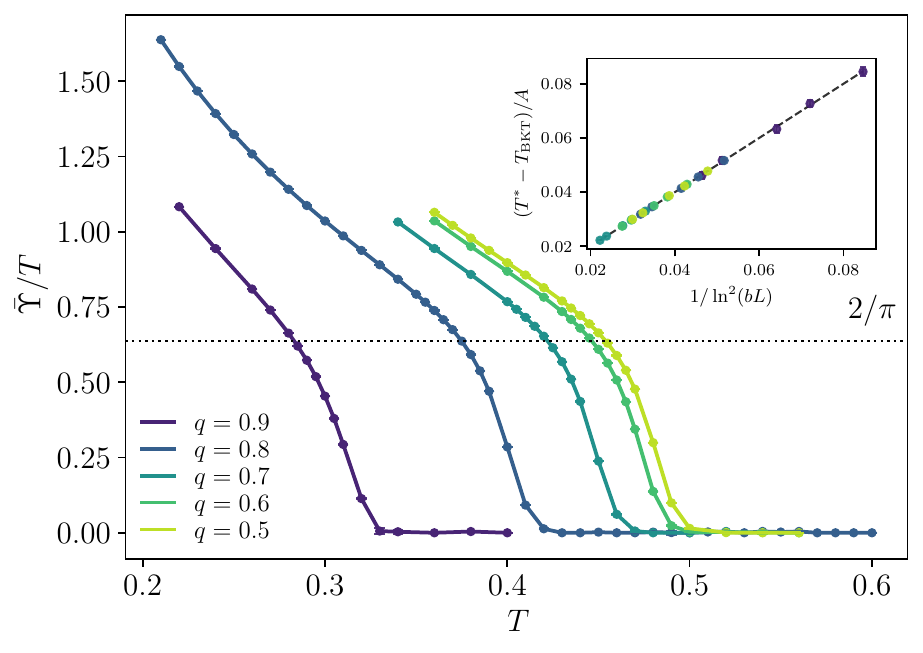}
 \caption{Geometric-mean helicity modulus divided by temperature, $\bar\Upsilon(L,T)/T$, for $L=200$. The horizontal dashed line marks the universal jump value $2/\pi$. Each curve crosses $2/\pi$ only once, consistent with a single BKT transition for each $q$. Inset: the $\left[T^{*}(L;q)-T_{\mathrm{BKT}}(q)\right]/A(q)$ versus $\ln^{-2}\left[b(q)L\right]$ data and the corresponding linear fits for different $q$ values as a quality check. }
 \label{fig:L200_helicity_modulus}
\end{figure}

For each $q$ and $L$, we compute $\Upsilon_x(L,T)$ and $\Upsilon_y(L,T)$ and form
\[
 \bar\Upsilon(L,T)=\sqrt{\Upsilon_x(L,T)\Upsilon_y(L,T)}.
\]
The size-dependent crossing temperature $T^*(L)$ is obtained from $\bar\Upsilon(L,T)/T=2/\pi$, and Eq.~(\ref{eq:BKT_finite_size}) is then used to fit $T_{\mathrm{BKT}}(q)$. As a quality check of the extrapolation, we also plot $\left[T^{*}(L;q)-T_{\mathrm{BKT}}(q)\right]/A(q)$ versus $\ln^{-2}\left[b(q)L\right]$ for different $q$ values and perform linear fits in the inset of Fig.~\ref{fig:L200_helicity_modulus}. The linearity with slope 1 of this plot confirms the validity of the finite-size form in Eq.~\eqref{eq:BKT_finite_size}. The resulting values are listed in Table~\ref{tab:BKT_transition_temperatures}.

\begin{table}[tb]
 \centering
 \begin{ruledtabular}
 \begin{tabular}{lccccc}
 $q$ & 0.5 & 0.6 & 0.7 & 0.8 & 0.9 \\
 \hline
 $T_{\mathrm{BKT}}$ & 0.4472(3) & 0.4392(4) & 0.4144(6) & 0.3705(5) & 0.2820(6) \\
 \end{tabular}
 \end{ruledtabular}
 \caption{Equilibrium BKT temperatures $T_{\mathrm{BKT}}$ of the anisotropic XY model. The temperatures are obtained from the finite-size extrapolation Eq.~\eqref{eq:BKT_finite_size} using $L=60,80,100,160$, and $200$, and are measured in units of $E_*$. Parentheses denote statistical fitting uncertainties. }
 \label{tab:BKT_transition_temperatures}
\end{table}

For the largest size $L=200$, Fig.~\ref{fig:L200_helicity_modulus} shows $\bar\Upsilon/T$ for all $q$. In each case there is only one crossing with $2/\pi$, consistent with a single BKT transition. The transition temperature decreases as the anisotropy increases, as expected from the reduction of the geometric-mean stiffness.

\subsection{Nonequilibrium}

Having determined the equilibrium BKT temperature, we now turn to the criterion-dependent transport-extracted temperatures obtained from RSJD simulations. We first keep the dissipative channel isotropic, $r_{0,x}=r_{0,y}=1$, so that the transport anisotropy comes only from $E_{J,x}\neq E_{J,y}$. We use $q=0.6$ as the representative example and then compare with $q=0.7$. After that, we turn on anisotropy in the dissipative channel. 

\subsubsection{Transport criteria: $q=0.6$ as an example}
\begin{figure}[tb]
 \centering
 \includegraphics[width=1.\linewidth]{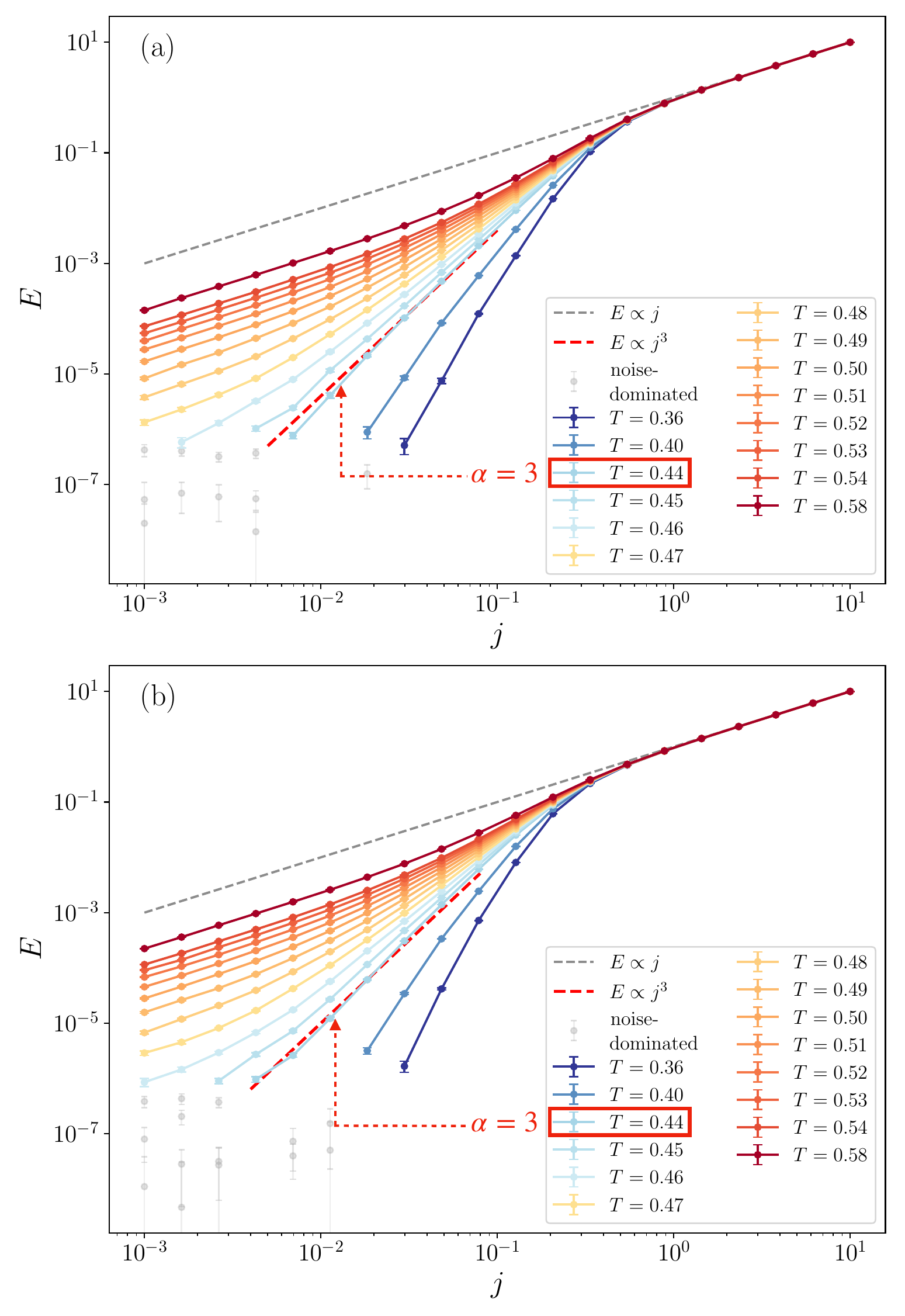}
 \caption{Dimensionless $E$--$j$ characteristics for $L=100$ and $q=0.6$ with current applied along (a) $x$ and (b) $y$. The dashed guide lines indicate Ohmic behavior, $E\propto j$, and the BKT critical power law, $E\propto j^3$. Gray data points are dominated by numerical noise and are excluded from the quantitative analysis.}
 \label{fig:64_IV_curves}
\end{figure}

The nonequilibrium analysis begins from the $E$--$j$ curves obtained by integrating Eqs.~(\ref{eq:RSJD_EoM_theta_dimensionless}) and (\ref{eq:RSJD_EoM_delta_dimensionless}) for currents applied along $x$ and $y$. The integration is performed by the second-order Runge-Kutta-Helfand-Greenside algorithm for stochastic dynamics\cite{helfand1979numerical,greenside1981numerical}. The time step $\Delta\tau$ is set to be $0.02$, and the long-time average in Eq.~\eqref{eq:electric_field_averaged} is calculated over $\sim10^7$ steps for large current densities and $\sim10^8$ steps for small current densities. Representative data for $L=100$ and $q=0.6$ are shown in Fig.~\ref{fig:64_IV_curves}. Gray points are discarded in the analysis since the signal is dominated by numerical noise. 

\paragraph{$\alpha=3$ criterion.}
A standard way to analyze nonlinear transport near a BKT transition is to define the local slope
\begin{equation}
 \alpha_\mu(T,j_\mu)\equiv\frac{\dd\ln E_\mu}{\dd\ln j_\mu}.
\end{equation}
In the ideal Ambegaokar--Halperin--Nelson--Siggia (AHNS) picture\cite{ambegaokar1978dissipation,ambegaokar1980dynamics,minnhagen1987two}, the transition is identified by $\alpha(T_{\mathrm{BKT}}^-)=3$. The temperature obtained from this condition will be treated as a critical-scaling transport-extracted temperature. 

For the finite anisotropy considered here, this criterion is unchanged by the rescaling used in Sec.~\ref{sec:model}: the stiffness entering AHNS is the renormalized geometric mean $\bar\Upsilon_R$, and a directional $E_\mu$ and $j_\mu$ acquire only constant metric factors. Thus a power law $E_\mu\propto j_\mu^\alpha$ keeps the same exponent. Since $\alpha(T)=1+\pi\bar\Upsilon_R(T)/T$ and $\bar\Upsilon_R(T_{\mathrm{BKT}}^-)=2T_{\mathrm{BKT}}/\pi$, both directions give $\alpha(T_{\mathrm{BKT}}^{-})=3$. 

\begin{figure}[tb]
 \centering
 \includegraphics[width=1.\linewidth]{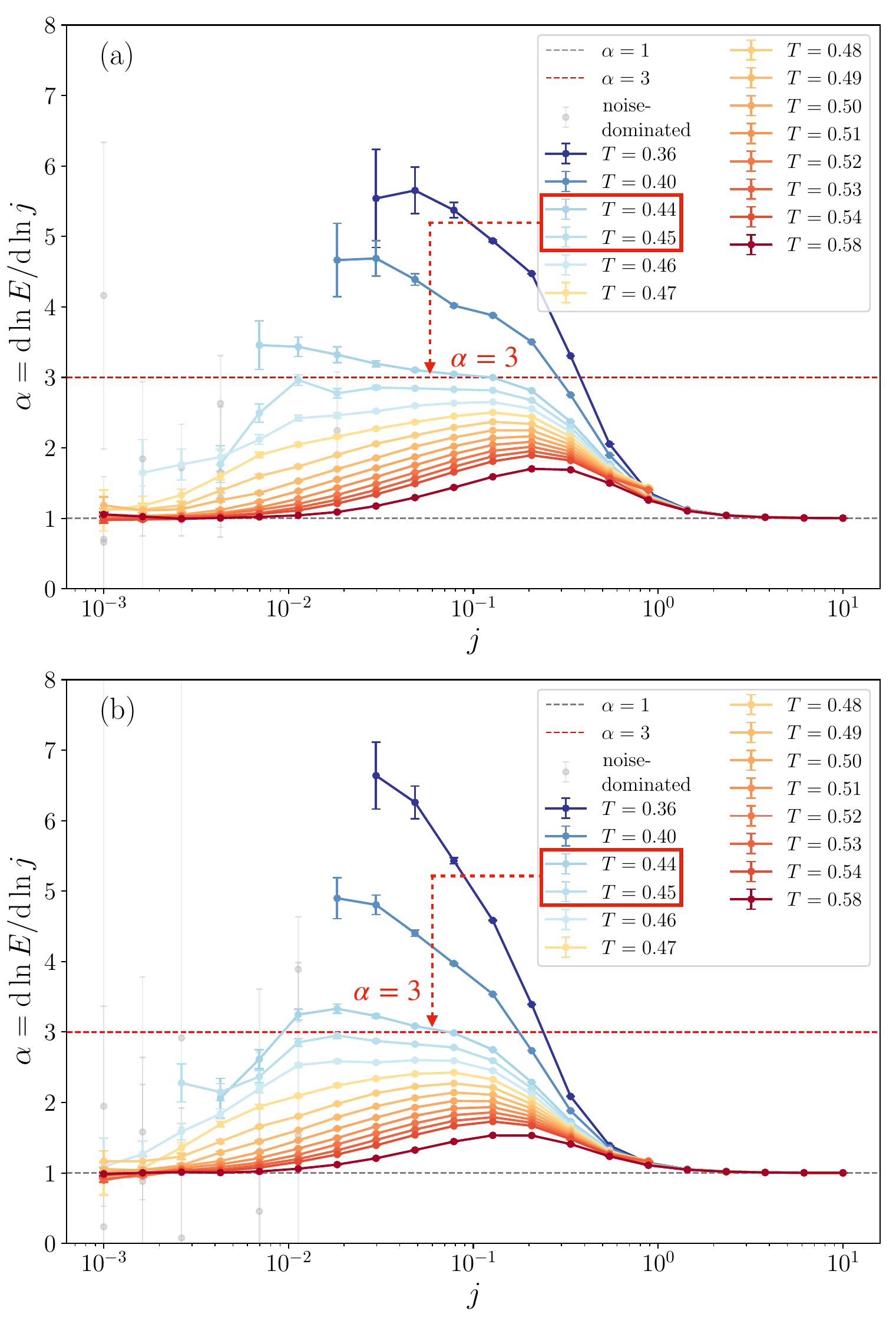}
 \caption{Local nonlinear transport exponent $\alpha_\mu(T,j_\mu)=(\dd\ln E_\mu)/(\dd\ln j_\mu)$ for $L=100$ and $q=0.6$, with currents applied along (a) $x$ and (b) $y$. The horizontal dashed line marks the BKT criterion $\alpha=3$. The curves from $T=0.44$ to $T=0.45$ bracket the crossing for both current directions, consistent with the equilibrium estimate $T_{\mathrm{BKT}}\approx 0.439$. }
 \label{fig:64_alphaj_curves}
\end{figure}

Figure~\ref{fig:64_alphaj_curves} shows the numerically differentiated $\alpha$--$j$ data. For both current directions, the $\alpha=3$ condition falls between $T=0.44$ and $0.45$, and with the present temperature resolution we take
\begin{equation}
 T_x^{\alpha=3}\approx T_y^{\alpha=3}\approx 0.44.
\end{equation}
This agrees well with the equilibrium estimate $T_{\mathrm{BKT}}\approx 0.439$ in Table~\ref{tab:BKT_transition_temperatures}. 

\paragraph{Dynamic finite-size scaling.}
We next examine dynamic finite-size scaling (FSS). In two dimensions, the Fisher--Fisher--Huse form can be written as\cite{fisher1991thermal,kim1999vortex,medvedyeva2001ubiquitous}
\begin{equation}\label{eq:FFH_revised}
 \frac{E_\mu}{j_\mu}=\xi^{-z}\chi_\pm\!\left(\frac{j_\mu\xi}{T}\right),
\end{equation}
where $\xi$ is the correlation length, $z$ is the dynamic critical exponent, and $\chi_{\pm}$ are the scaling functions above and below the BKT transition temperature. In practice, we keep the conventional value $z=2$ fixed to find the FSS transport-extracted temperature where the data collapse. A finite spatial anisotropy only changes the metric of the correlation lengths, i.e., $\xi_\mu=c_\mu\xi'$. Thus, the relaxation time still scales as $\tau\sim(\xi')^z$ with one exponent. At the transition the divergence of $\xi$ is cut off by the system size, so one expects
\begin{equation}\label{eq:FFH_FSS_revised}
 \frac{E_\mu}{j_\mu}L^z=\mathcal \chi_\pm\!\left(\frac{j_\mu L}{T}\right).
\end{equation}

\begin{figure}[tb]
 \centering
 \includegraphics[width=1.\linewidth]{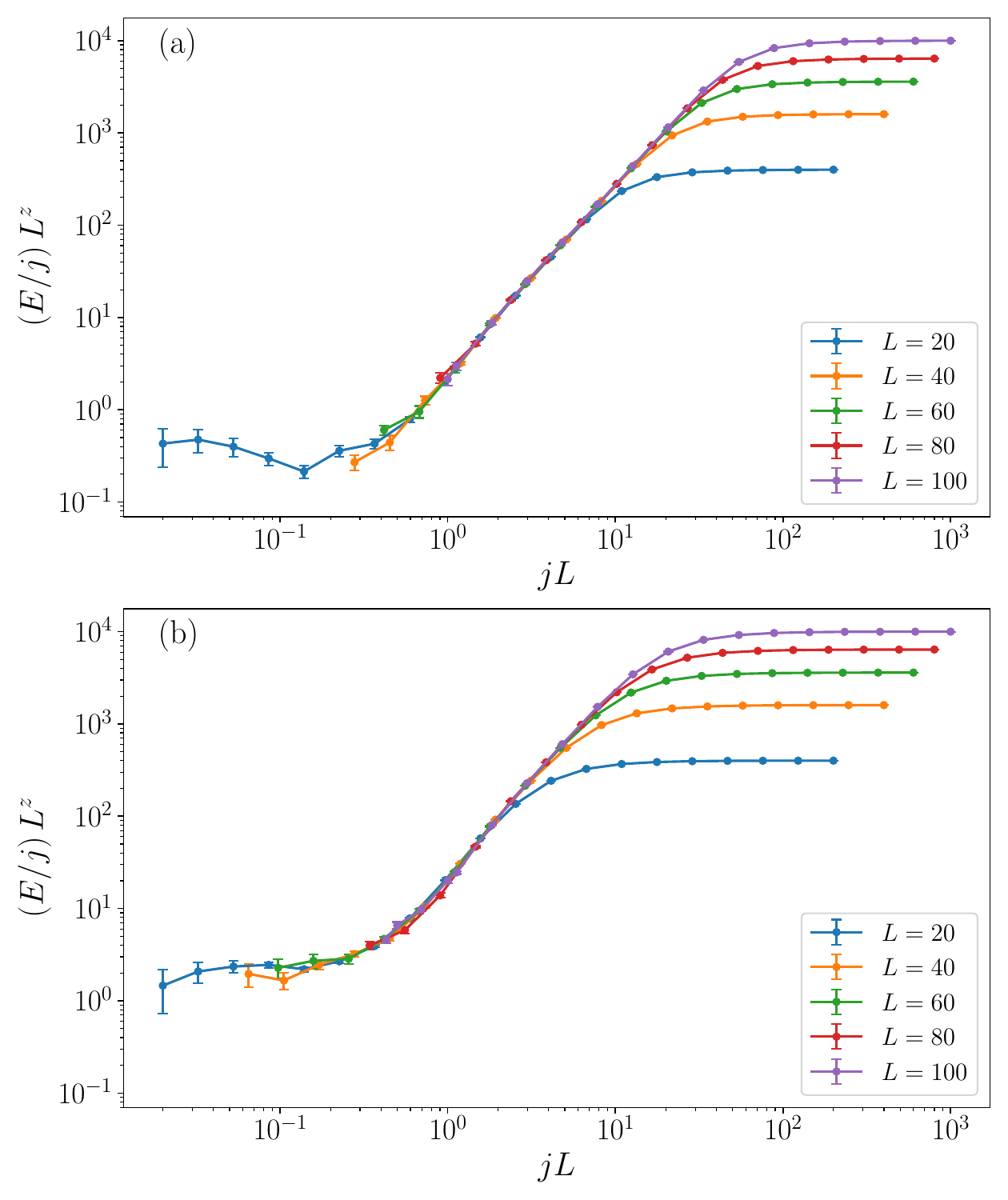}
 \caption{Dynamic finite-size scaling of the $E$--$j$ data for $q=0.6$ at $T=0.44$. The scaled resistance $(E_\mu/j_\mu)L^z$ is plotted against $j_\mu L$ for (a) $j\parallel x$ and (b) $j\parallel y$. System sizes are $L=20,40,60,80,100$, and the collapse is obtained with $z=2$. The same temperature and exponent give comparable collapses for both current directions. }
 \label{fig:64_FFH_scaling}
\end{figure}

Using $L=20$, $40$, $60$, $80$, and $100$, we plot the finite-size scaling form at $T=0.44$, the temperature grid point closest to the equilibrium estimate of $T_{\mathrm{BKT}}$, with the conventional dynamic exponent $z=2$ \cite{hohenberg1977theory,kim1999vortex,medvedyeva2001ubiquitous}. As shown in Fig.~\ref{fig:64_FFH_scaling}, the same choice of $T$ and $z$ gives comparable collapses for currents driven along $x$ and $y$: 
\begin{equation}
 T_x^{\mathrm{FSS}}\approx T_y^{\mathrm{FSS}}\approx 0.44.
\end{equation}
Like the $\alpha=3$ criterion, dynamic FSS successfully accounts for the system cutoffs and yields a single critical-scaling transport-extracted temperature for probe currents along both orthogonal directions, and does not indicate a resolvable directional splitting. 

\begin{figure}[tb]
 \centering
 \includegraphics[width=1.\linewidth]{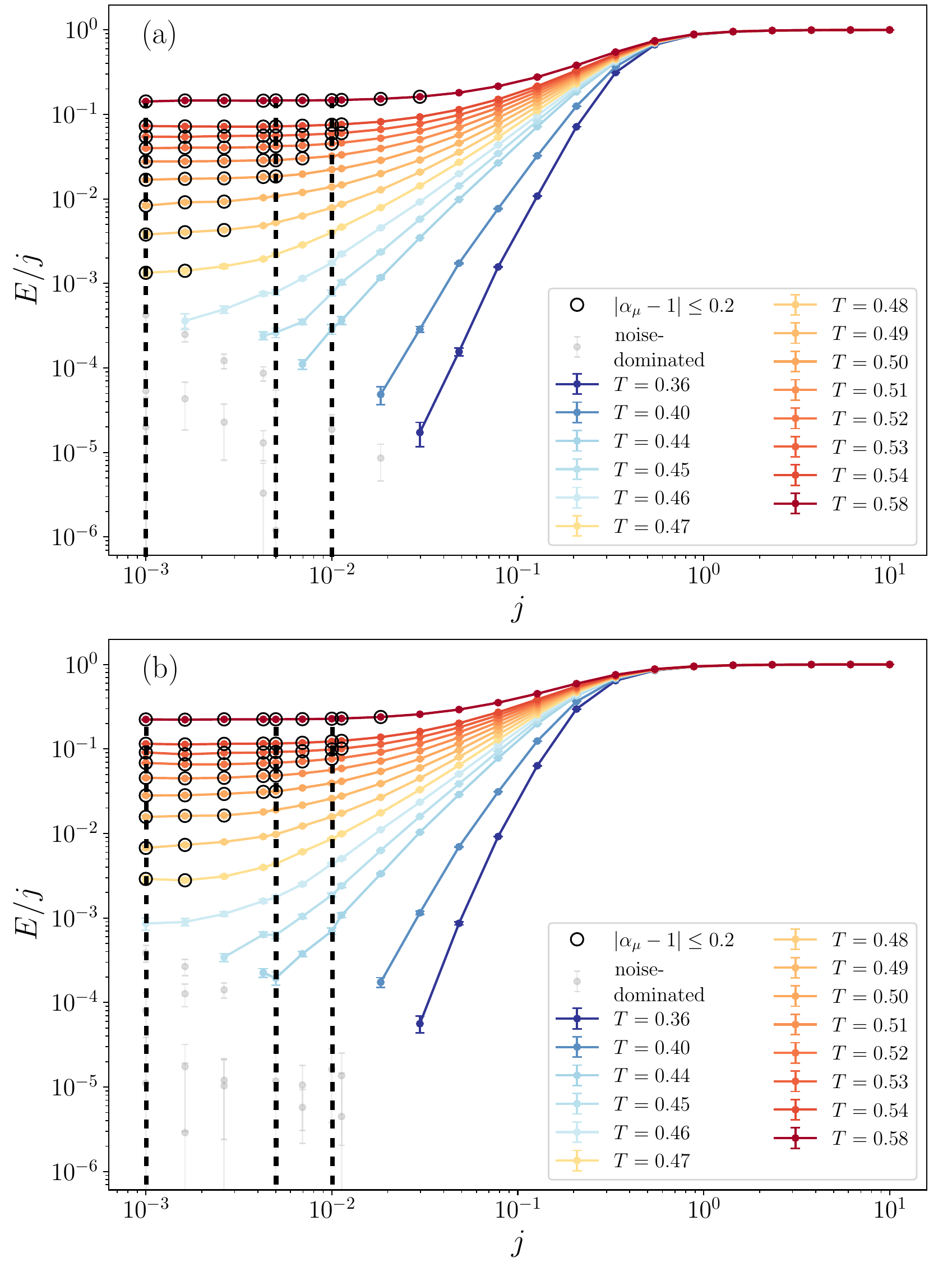}
 \caption{Linear-response check for the probe currents used to construct $r_{\mathrm{lin},\mu}$. The ratio $E_\mu/j_\mu$ is plotted as a function of $j_\mu$ for $L=100$ and $q=0.6$, with current along (a) $x$ and (b) $y$. The vertical black dashed lines mark $j=0.01$, $0.005$, and $0.001$. Black open circles denote data points retained as linear-response data, selected by the criterion $|\alpha_\mu-1|\leq 0.2$.}
 \label{fig:64_linear_regime}
\end{figure}

\paragraph{Linear-response window and $r_{\mathrm{lin}}$--$T$ curves.}
 To construct an operational linear-resistance curve from the finite-current transport data, one must identify a current window in which the response is approximately Ohmic. We define the dimensionless resistance at a given probe current as
\begin{equation}
 r_{\mathrm{lin},\mu}=\frac{E_\mu}{j_\mu},
\end{equation}
 for which the low-current response exhibits a plateau consistent with $\alpha_\mu\simeq 1$. In practice, we use $|\alpha_{\mu}-1| \leq 0.2$ as the criterion for assigning a data point to the linear-response window\footnote{ Using a narrower tolerance gives fewer accepted data points near the low-resistance side of the transition, whereas a broader tolerance gives a larger operational window. The qualitative comparison between the two current directions is based on the anisotropic shape of the underlying transport curves, rather than the special value of this tolerance. }. Figure~\ref{fig:64_linear_regime} illustrates this for $j=0.01$, $0.005$, and $0.001$. 

\begin{figure}[tb]
 \centering
 \includegraphics[width=1.\linewidth]{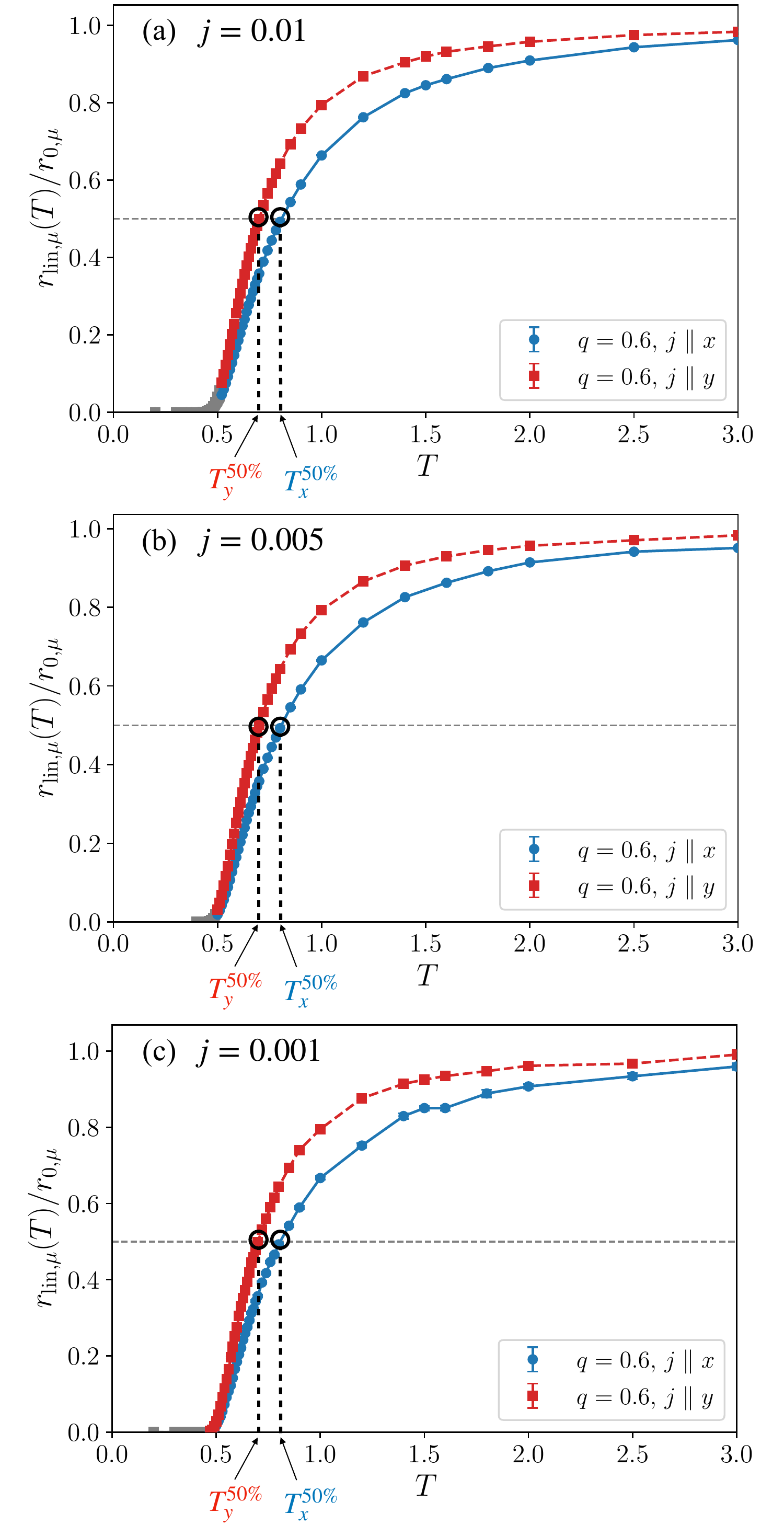}
 \caption{Normalized linear resistance $r_{\mathrm{lin},\mu}/r_{0,\mu}$ for $L=100$ and $q=0.6$ with isotropic dissipation $r_{0,x}=r_{0,y}=1$ at probe currents (a) $j=0.01$, (b) $j=0.005$, and (c) $j=0.001$. Blue circles and red squares denote currents applied along $x$ and $y$, respectively. The horizontal dashed line marks the empirical $50\%$ threshold, $r_{\mathrm{lin},\mu}/r_{0,\mu}=0.5$. Gray points are outside the verified linear-response window and are not used for extracting threshold temperatures. }
 \label{fig:64_rt}
\end{figure}

Choosing an isotropic dissipative channel ($r_{0,x}=r_{0,y}=1$), the corresponding $r_{\mathrm{lin}}$--$T$ curves are shown in Fig.~\ref{fig:64_rt}. We use two empirical threshold criteria common in experimental analysis: 
\begin{equation}
 \frac{r_{\mathrm{lin},\mu}}{r_{0,\mu}}=0.5,
 \qquad
 \frac{r_{\mathrm{lin},\mu}}{r_{0,\mu}}=0.01,
\end{equation}
yielding threshold transport-extracted temperatures $T_\mu^{50\%}$ and $T_\mu^{1\%}$, respectively. 

For the $50\%$ criterion we obtain
\[
\begin{aligned}
 j=0.01:&\qquad T_x^{50\%}=0.808, \quad T_y^{50\%}=0.700,\\
 j=0.005:&\qquad T_x^{50\%}=0.807, \quad T_y^{50\%}=0.700,\\
 j=0.001:&\qquad T_x^{50\%}=0.808, \quad T_y^{50\%}=0.701.
\end{aligned}
\]

These temperatures are obviously direction dependent, but they are also far above the $T_{\mathrm{BKT}}$. Therefore, $T_{\mu}^{50\%}$ should be viewed as a threshold transport-extracted temperature that diagnoses the direction-dependent reshaping of the resistive curves, rather than as an estimate of the equilibrium BKT temperature.

\begin{figure}[tb]
 \centering
 \includegraphics[width=1.\linewidth]{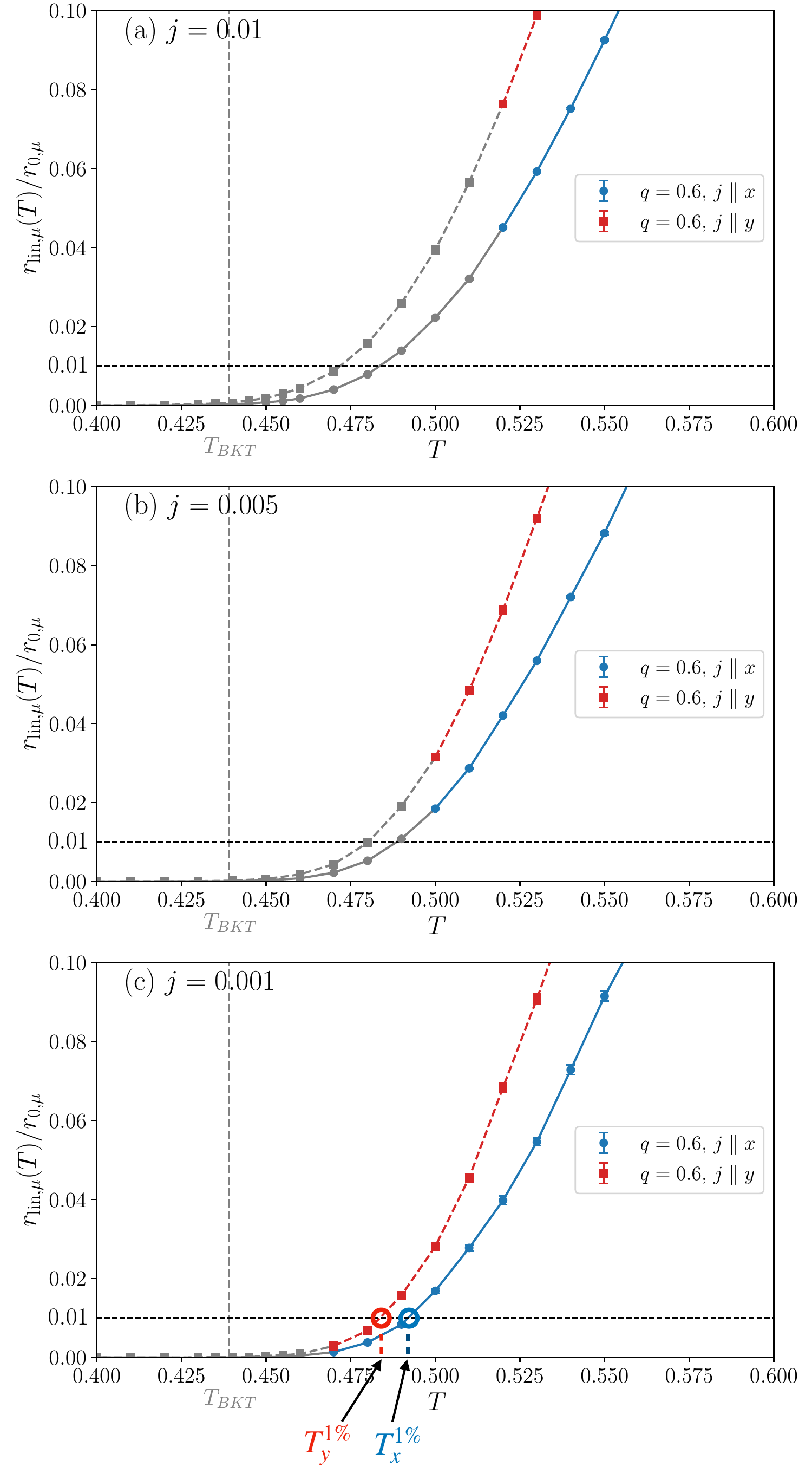}
 \caption{Zoom-in of low-resistance regions of the normalized linear resistance $r_{\mathrm{lin},\mu}/r_{0,\mu}$ for $L=100$, $q=0.6$, and isotropic dissipation $r_{0,x}=r_{0,y}=1$ at (a) $j=0.01$, (b) $j=0.005$, and (c) $j=0.001$. The horizontal dashed line marks the $1\%$ threshold $r_{\mathrm{lin},\mu}/r_{0,\mu}=0.01$. The vertical dashed line indicates the equilibrium $T_{\mathrm{BKT}}\approx 0.439$. Gray points lie outside the verified linear-response window and are not used for extracting threshold temperatures.}
 \label{fig:64_rt_zoomin}
\end{figure}

A lower resistance threshold moves the extracted temperature closer to $T_{\mathrm{BKT}}$, but only if the intersection still lies inside the linear-response window. Figure~\ref{fig:64_rt_zoomin} shows the relevant low-resistance region. For $j=0.01$ and $0.005$, the $1\%$ threshold is reached only below the lowest temperature at which the probe current remains linear, so no reliable $T_\mu^{1\%}$ can be assigned. For $j=0.001$, we can extract
\[
 T_x^{1\%}=0.492,
 \qquad
 T_y^{1\%}=0.484.
\]
Compared with $T_{\mu}^{50\%}$, these values are much closer to $T_{\mathrm{BKT}}$, and they show a small directional splitting. As the temperature is lowered, $r_{\mathrm{lin},x}$ obtained from the $x$--direction with larger $E_{J,x}$ will drop below the $1\%$ threshold earlier than the $y$--direction, giving $T_x^{1\%}>T_y^{1\%}$. 

\begin{figure}[tb]
 \centering
 \includegraphics[width=1.\linewidth]{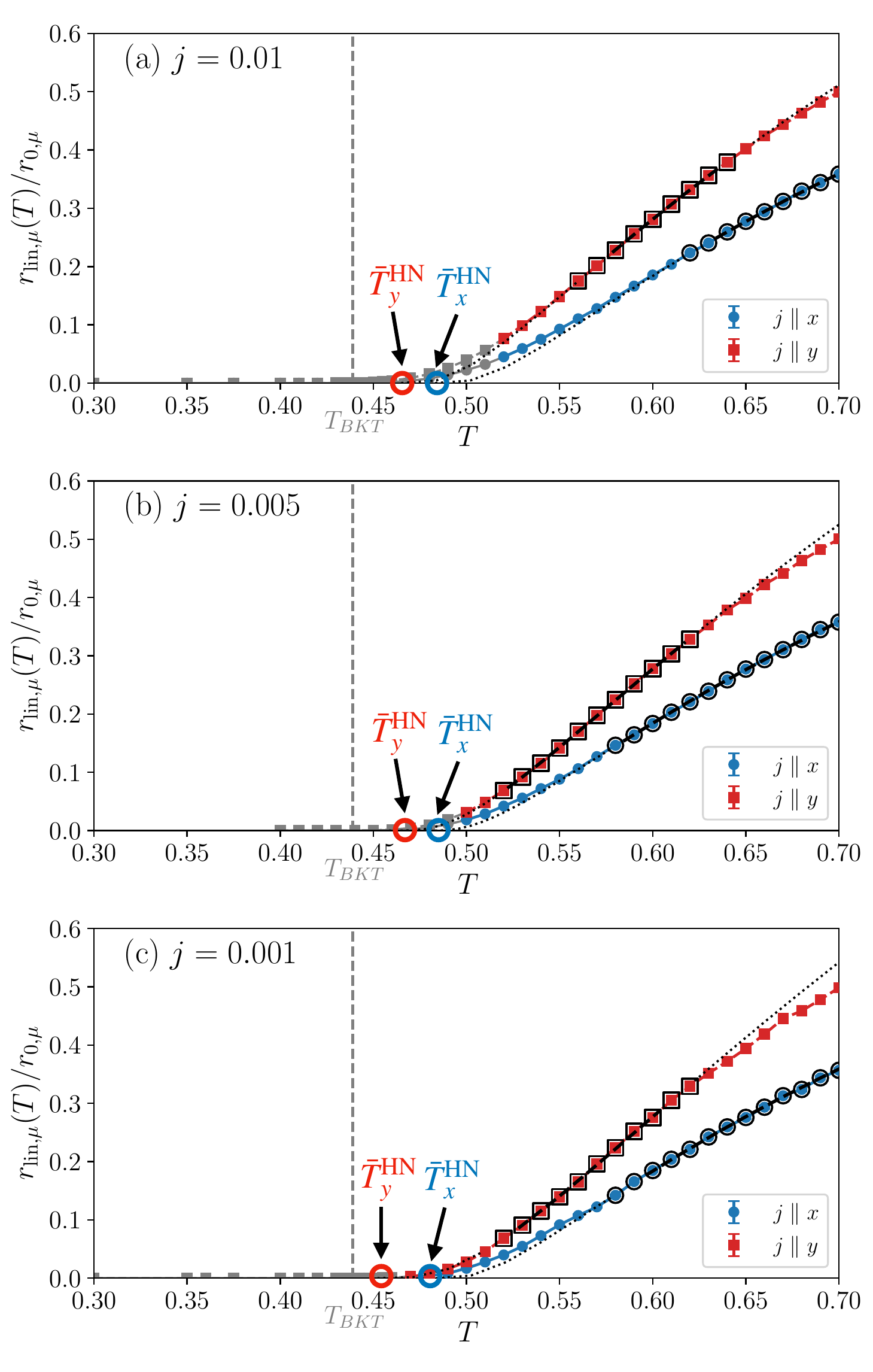}
 \caption{Halperin--Nelson fits of the normalized linear resistance for $L=100$, $q=0.6$, and isotropic dissipation $r_{0,x}=r_{0,y}=1$ at (a) $j=0.01$, (b) $j=0.005$, and (c) $j=0.001$. Blue circles and red squares denote currents along $x$ and $y$, respectively. Dashed black curves are fits to Eq.~\eqref{eq:HN_fit} over the representative windows marked by black open symbols. The vertical gray dashed line indicates the equilibrium $T_{\mathrm{BKT}}\approx 0.439$. Gray points lie outside the verified linear-response window and are not used for the HN fitting. The fitted values satisfy $T_x^{\mathrm{HN}}>T_y^{\mathrm{HN}}$ even though the critical-scaling criteria give a single transition temperature.}
 \label{fig:64_hn_fit}
\end{figure}

\paragraph{Halperin--Nelson fits.}
Finally, we fit the linear resistance above the transition to the Halperin--Nelson form \cite{halperin1979resistive},
\begin{equation}\label{eq:HN_fit}
 r_{\mathrm{lin},\mu}(T)=r_{\mu}^{\mathrm{HN},0}
 \exp\!\left[-b_\mu\left(\frac{T_\mu^{\mathrm{HN}}}{T-T_\mu^{\mathrm{HN}}}\right)^{1/2}\right].
\end{equation}
Again, the fit is performed only on points inside the linear regime. Representative fits are shown in Fig.~\ref{fig:64_hn_fit}. We emphasize that $T_\mu^{\mathrm{HN}}$ is used here as an HN transport-extracted temperature obtained from a finite-size, finite-current resistance curve, rather than as an independent estimate of the thermodynamic $T_{\mathrm{BKT}}$. To estimate the sensitivity to the chosen fitting interval, we start from the manually selected central window indicated by the black open symbols in Fig.~\ref{fig:64_hn_fit} and vary the interval by one or two temperature points around the central window\footnote{If the central window contains the temperature interval $[T_{\mathrm{left}},T_{\mathrm{right}}]$, we generate shifted windows by moving the left and right endpoints independently by $0,\pm 1,\pm 2$ temperature-grid points, while retaining only successful fits with positive resistance data and with all points inside the linear-response regime. This procedure gives up to $25$ window variants for each current direction. }. The resulting representative fitting parameters for $q=0.6$ and isotropic dissipation are summarized in Table~\ref{tab:HN_q06_details}. Unlike the critical-scaling transport-extracted temperatures obtained by the $\alpha=3$ and FSS criteria, the HN transport-extracted temperatures show a directional separation, $T_x^{\mathrm{HN}}>T_y^{\mathrm{HN}}$. 

\begin{table}[tb]
 \caption{
 Representative Halperin--Nelson fitting parameters for the normalized linear resistance at $L=100$, $q=0.6$, and isotropic dissipation $r_{0,x}=r_{0,y}=1$. The HN fit uses Eq.~\eqref{eq:HN_fit} over the representative central window shown in Fig.~\ref{fig:64_hn_fit}, with the fit performed in log space. $T_\mu^{\mathrm{HN}}$ is reported as the mean over shifted-window variants; parentheses denote the combined uncertainty. The parameters $r_{\mu}^{\mathrm{HN},0}$, $b_\mu$, and $\chi_\nu^2$ refer to the central-window fit.
 }
 \label{tab:HN_q06_details}
 \centering
 \begin{ruledtabular}
 \begin{tabular}{cccccc}
 $j$ & Current direction & $T_\mu^{\mathrm{HN}}$ & $r_{\mu}^{\mathrm{HN},0}$ & $b_\mu$ & $\chi_\nu^2$ \\
 \hline
 $0.01$ & $j\parallel x$ & $0.483(6)$ & 2.286 & 1.240 & 1.89 \\
 $0.01$ & $j\parallel y$ & $0.465(5)$ & 3.341 & 1.335 & 1.80 \\
 $0.005$ & $j\parallel x$ & $0.477(5)$ & 2.450 & 1.316 & 2.17 \\
 $0.005$ & $j\parallel y$ & $0.458(6)$ & 4.230 & 1.516 & 4.45 \\
 $0.001$ & $j\parallel x$ & $0.484(6)$ & 2.264 & 1.232 & 0.58 \\
 $0.001$ & $j\parallel y$ & $0.452(6)$ & 5.300 & 1.688 & 0.71 \\
 \end{tabular}
 \end{ruledtabular}
\end{table}

 The values reported in Table~\ref{tab:HN_q06_details} use the mean of $T_\mu^{\mathrm{HN}}$ over the shifted-window variants. The quoted uncertainty combines the statistical uncertainty of the central nonlinear fit and the systematic window uncertainty $\sigma_{T,\mu}^{\mathrm{HN}}=\sqrt{\left(\sigma_{\mathrm{stat},\mu}^{\mathrm{HN}}\right)^2+\left(\sigma_{\mathrm{win},\mu}^{\mathrm{HN}}\right)^2}$, where $\sigma_{\mathrm{stat},\mu}^{\mathrm{HN}}$ is the contribution from the central nonlinear fit, and $\sigma_{\mathrm{win},\mu}^{\mathrm{HN}}$ is the standard deviation of $T_\mu^{\mathrm{HN}}$ over the shifted-window variants. The reduced $\chi^2_\nu$ in Table~\ref{tab:HN_q06_details} is reported for the central-window log-space fit. 

The physical origin of this apparent splitting lies in the distinction between the true asymptotic critical regime and the operational crossover regime. However, in any numerical simulation and indeed in any real experiment, finite system size, finite probe current, and the available linear-response window cut off this divergence and restrict the fit to a higher-temperature crossover regime\cite{simkin1997finite}. Moreover, Ref.~\cite{benfatto2009broadening} shows that finite-size effects and inhomogeneity can broaden the resistive transition, generate resistive tails, and make the physical interpretation of HN fitting parameters nontrivial. In the present clean model, the cutoff is caused by the finite $L$, finite $j$, and finite fitting-window constraints of the transport simulation. Consequently, the linear-response $r_{\mathrm{lin}}$--$T$ data are necessarily truncated, forcing the HN fits to be performed in a slightly higher-temperature crossover regime (as evidenced by $T^{\mathrm{HN}}_{\mu} > T_{\mathrm{BKT}}$ in all our fits). In this crossover regime, non-universal short-range physics---such as the bare anisotropic Josephson couplings---dominates the shape of the resistive curve. Because these short-range parameters are anisotropic, the resistive curves are reshaped differently along $x$ and $y$, causing the HN fit to yield two distinct apparent HN transport-extracted temperatures $T_{\mu}^{\mathrm{HN}}$'s even though the thermodynamic transition remains single.

In the combined limit $L\to\infty$ and $j\to 0$, the finite-size and finite-current cutoffs are removed and the HN fit can access the asymptotic critical regime. In that limit, both $T_x^{\mathrm{HN}}$ and $T_y^{\mathrm{HN}}$ are expected to converge to the single thermodynamic $T_{\mathrm{BKT}}$ dictated by the geometric-mean stiffness. The directional separation observed at the accessible $(L,j)$ is therefore a finite-size, finite-current feature of the clean anisotropic baseline, not a property of the thermodynamic transition.

The \(q=0.6\) example already contains the main distinction. Critical-scaling transport criteria ($\alpha=3$ and FSS) give a single transport-extracted temperature near $0.44$, consistent with the equilibrium BKT temperature. By contrast, curve-shape transport criteria applied to the low-current $r_{\mathrm{lin}}$--$T$ curves yield direction-dependent transport-extracted temperatures, with the larger $E_{J,\mu}$ direction giving the higher extracted value. Within the present model, these curve-shape transport-extracted temperatures are operational quantities rather than evidence for a split thermodynamic transition.

\subsubsection{Varying the anisotropy parameter $q$}
We now keep the dissipative channel isotropic, $r_{0,x}=r_{0,y}=1$, and compare $q=0.6$ with $q=0.7$ on the same $L=100$ lattice, using $q=0.5$ as the isotropic reference. The critical-scaling criteria are collected in Table~\ref{tab:iso_64_73_1}, while the curve-shape-based temperatures extracted from the $r_{\mathrm{lin}}$--$T$ curves are collected in Table~\ref{tab:iso_64_73_2}.

\begin{table}[tb]
 \caption{Dimensionless critical-scaling transport-extracted temperatures obtained from critical-scaling criteria with isotropic shunt resistance $r_{0,x}=r_{0,y}=1$. The $\alpha=3$ criterion uses $E$--$j$ data for $L=100$, while the dynamic FSS analysis uses $L=20,40,60,80,100$ with $z=2$. }
 \label{tab:iso_64_73_1}
 \centering
 \begin{ruledtabular}
 \begin{tabular}{lcccc}
 Criterion & Current direction & $q=0.5$ & $q=0.6$ & $q=0.7$ \\
 \hline
 $\alpha=3$ & $j\parallel x$ & 0.45 & 0.44 & 0.42 \\
 $\alpha=3$ & $j\parallel y$ & 0.45 & 0.44 & 0.42 \\
 dynamic FSS & $j\parallel x$ & 0.45 & 0.44 & 0.42 \\
 dynamic FSS & $j\parallel y$ & 0.45 & 0.44 & 0.42 \\
 \end{tabular}
 \end{ruledtabular}
\end{table}

\begin{table}[tb]
 \caption{Dimensionless curve-shape transport-extracted temperatures obtained from the $r_{\mathrm{lin}}$--$T$ curves for $L=100$ with isotropic shunt resistance $r_{0,x}=r_{0,y}=1$. The HN entries are obtained by fitting Eq.~\eqref{eq:HN_fit} within the verified linear-response window. The $1\%$ criterion means the $1\%$ threshold defined by $r_{\mathrm{lin},\mu}/r_{0,\mu}=0.01$ and is reported only for $j=0.001$, for which the threshold crossing lies inside the linear-response window. Parentheses for the HN entries denote the combined uncertainty from the covariance of the central-window nonlinear fit and the standard deviation over shifted fitting windows, and threshold values are direct interpolation read-offs. Dashes for $q=0.5$ indicate that no reliable threshold temperature is assigned within the verified linear-response window.}
 \label{tab:iso_64_73_2}
 \centering
 \begin{ruledtabular}
 \begin{tabular}{lccccc}
 Criterion & $j$ & Current direction & $q=0.5$ & $q=0.6$ & $q=0.7$ \\
 \hline
 HN fit & $0.01$ & $j\parallel x$ & 0.462(4) & 0.483(6) & 0.456(6) \\
 HN fit & $0.01$ & $j\parallel y$ & 0.462(4) & 0.465(5) & 0.430(10) \\
 HN fit & $0.005$ & $j\parallel x$ & 0.461(4) & 0.477(5) & 0.464(9) \\
 HN fit & $0.005$ & $j\parallel y$ & 0.461(4) & 0.458(6) & 0.434(9) \\
 HN fit & $0.001$ & $j\parallel x$ & 0.460(6) & 0.484(6) & 0.448(13) \\
 HN fit & $0.001$ & $j\parallel y$ & 0.460(6) & 0.452(6) & 0.438(11) \\
 $1\%$ & $0.001$ & $j\parallel x$ & \textemdash & 0.492 & 0.472 \\
 $1\%$ & $0.001$ & $j\parallel y$ & \textemdash & 0.484 & 0.456 \\
 \end{tabular}
 \end{ruledtabular}
\end{table}

The $\alpha=3$ and FSS criteria again select a single temperature for each $q$. By contrast, the stronger splitting at $q=0.7$ is visible in Fig.~\ref{fig:comparert}(c), where the low-resistance threshold is crossed at more clearly separated temperatures than in Fig.~\ref{fig:comparert}(b). 
\begin{equation}\label{eq:1percent_isoR}
 \left.\frac{T_y^{1\%}}{T_x^{1\%}}\right|_{q=0.6}=0.984,
 \qquad
 \left.\frac{T_y^{1\%}}{T_x^{1\%}}\right|_{q=0.7}=0.966.
\end{equation}
Within this minimal model, increasing the coupling anisotropy enhances the apparent directional separation extracted from the $1\%$ threshold criterion, while the critical-scaling criteria continue to give a single transition. 

\subsubsection{Turning on anisotropy in the dissipative channel}
We next introduce anisotropy in the dissipative channel by setting
\[
 r_{0,x}=1,
 \qquad
 r_{0,y}=\frac12,
\]
and study $q=0.5$, $0.6$, and $0.7$ on the same $L=100$ lattice. Because $r_{0,\mu}$ enters only the RSJD dynamics, based on the analysis in Sec.~\ref{subsection:equilibrium}, this change does not modify the equilibrium transition. However, it can further reshape the nonequilibrium transport properties. 

While the phase-only model does not explicitly include microscopic single-electron physics (such as Fermi surface anisotropies), the anisotropic shunt resistance $r_{0,\mu}$ phenomenologically captures an anisotropic viscous drag (or mobility) for moving vortices. When a vortex moves across the system, the local phase winding generates a voltage $V \propto \dot{\phi}$, and the energy dissipated is proportional to $V^2 / r_{0,\mu}$. Therefore, making $r_{0,x} \neq r_{0,y}$ imposes an anisotropic vortex viscosity: vortices experience more friction moving along one axis than the other, which splits the simulated $r_{\mathrm{lin}}$--$T$ curves even if the underlying thermodynamic stiffness is perfectly isotropic.

At the level of nonlinear transport scaling, Table~\ref{tab:anisR_scaling} shows that the $\alpha=3$ and FSS criteria still give the same temperature for $j\parallel x$ and $j\parallel y$ for all three values of $q$.

\begin{table}[tb]
 \caption{Dimensionless transport temperatures extracted from critical-scaling criteria with anisotropic shunt resistance $r_{0,x}=1$ and $r_{0,y}=1/2$. The $\alpha=3$ criterion uses $E$--$j$ data for $L=100$, while the dynamic FSS analysis uses $L=20,40,60,80,100$ with $z=2$. }
 \label{tab:anisR_scaling}
 \centering
 \begin{ruledtabular}
 \begin{tabular}{lcccc}
 Criterion & Current direction & $q=0.5$ & $q=0.6$ & $q=0.7$ \\
 \hline
 $\alpha=3$ & $j\parallel x$ & 0.45 & 0.44 & 0.42 \\
 $\alpha=3$ & $j\parallel y$ & 0.45 & 0.44 & 0.42 \\
 dynamic FSS & $j\parallel x$ & 0.45 & 0.44 & 0.42 \\
 dynamic FSS & $j\parallel y$ & 0.45 & 0.44 & 0.42 \\
 \end{tabular}
 \end{ruledtabular}
\end{table}

For the curve-shape analysis, we use the same normalization convention as above and compare $r_{\mathrm{lin},\mu}/r_{0,\mu}$ between the two current directions. The HN temperatures and the $1\%$ threshold temperatures are summarized in Table~\ref{tab:anisR_RT}.

\begin{table}[tb]
 \caption{Dimensionless transport temperatures extracted from the $r_{\mathrm{lin}}$--$T$ curves for $L=100$ with anisotropic shunt resistance $r_{0,x}=1$ and $r_{0,y}=1/2$. The HN entries are obtained by fitting Eq.~\eqref{eq:HN_fit} within the verified linear-response window. The $1\%$ threshold is defined by $r_{\mathrm{lin},\mu}/r_{0,\mu}=0.01$. Parentheses for the HN entries denote the combined uncertainty from the covariance of the central-window nonlinear fit and the standard deviation over shifted fitting windows, and threshold values are direct interpolation read-offs.}
 \label{tab:anisR_RT}
 \centering
 \begin{ruledtabular}
 \begin{tabular}{lccccc}
 Criterion & $j$ & Current direction & $q=0.5$ & $q=0.6$ & $q=0.7$ \\
 \hline
 HN fit & $0.01$ & $j\parallel x$ & 0.492(4) & 0.477(3) & 0.444(5) \\
 HN fit & $0.01$ & $j\parallel y$ & 0.473(4) & 0.473(6) & 0.429(7) \\
 HN fit & $0.005$ & $j\parallel x$ & 0.484(3) & 0.470(3) & 0.445(4) \\
 HN fit & $0.005$ & $j\parallel y$ & 0.464(4) & 0.463(5) & 0.434(10) \\
 HN fit & $0.001$ & $j\parallel x$ & 0.488(4) & 0.472(8) & 0.443(15) \\
 HN fit & $0.001$ & $j\parallel y$ & 0.469(6) & 0.468(6) & 0.426(12) \\
 $1\%$ & $0.001$ & $j\parallel x$ & 0.500 & 0.496 & 0.478 \\
 $1\%$ & $0.001$ & $j\parallel y$ & 0.492 & 0.482 & 0.453 \\
 \end{tabular}
 \end{ruledtabular}
\end{table}

\begin{figure*}[tb]
 \centering
 \includegraphics[width=1.0\linewidth]{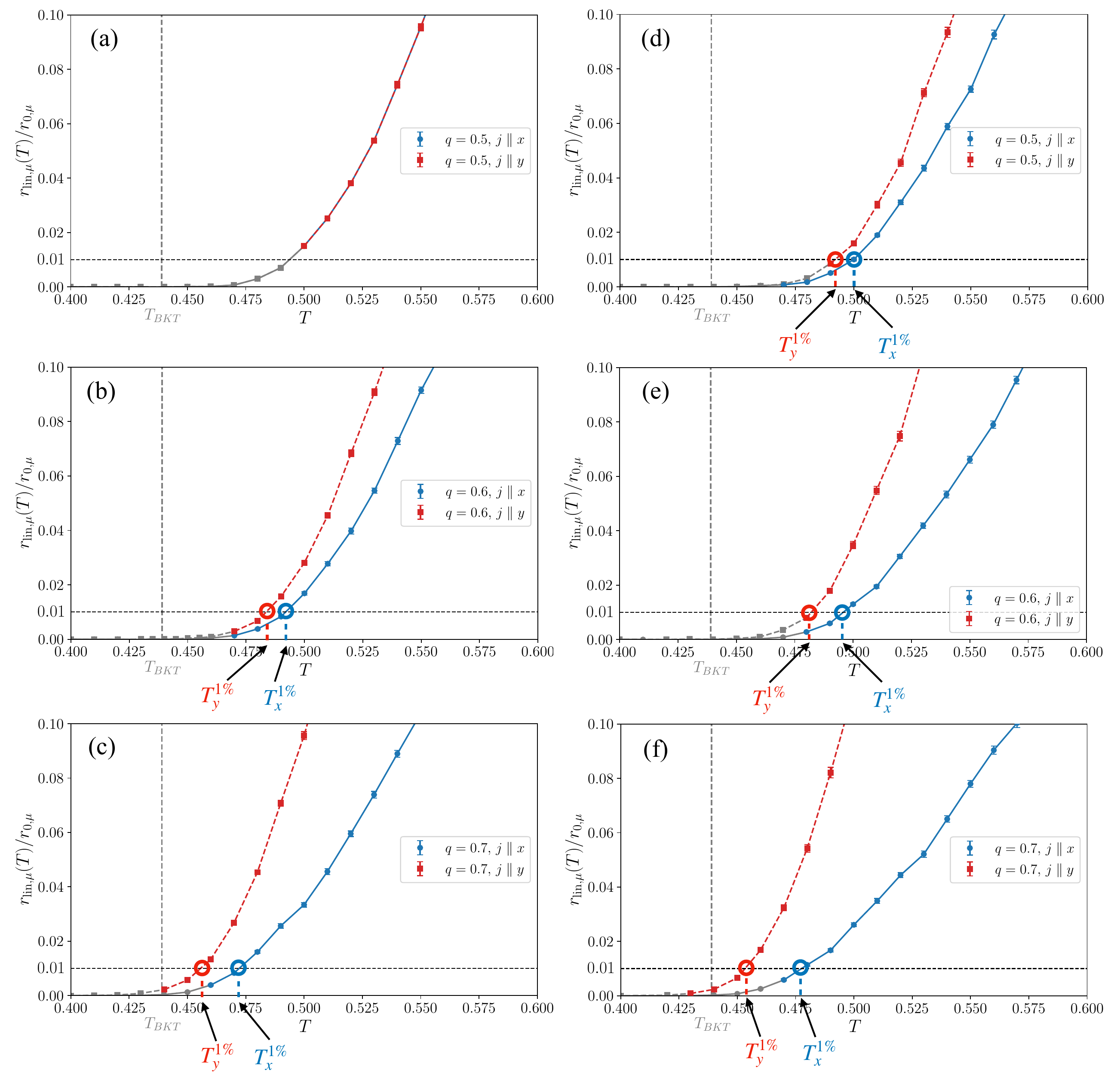}
 \caption{Effect of coupling and dissipation anisotropy on the low-resistance threshold. All panels show $r_{\mathrm{lin},\mu}/r_{0,\mu}$ versus $T$ for $L=100$ and $j=0.001$. The top row, (a)--(c), uses isotropic dissipation $r_{0,x}=r_{0,y}=1$ for $q=0.5,0.6,0.7$. The bottom row, (d)--(f), uses anisotropic dissipation $r_{0,x}=1$, $r_{0,y}=1/2$ for the same values of $q$. The horizontal dashed line marks the $1\%$ threshold $r_{\mathrm{lin},\mu}/r_{0,\mu}=0.01$, and the vertical dashed line marks the corresponding equilibrium $T_{\mathrm{BKT}}$. For $q=0.5$ with isotropic dissipation (panel (a)), no reliable threshold temperature is assigned within the verified linear-response window. However, with anisotropic dissipation (panel (d)), the $1\%$ threshold is crossed at different temperatures for the two directions, even though the underlying thermodynamic transition is isotropic. For $q=0.6$ and $q=0.7$ (panel (b), (c), (e) and (f)), assigning the larger shunt resistance to the more strongly coupled direction enhances the apparent threshold splitting, while the thermodynamic transition remains single.}
 \label{fig:comparert}
\end{figure*}

Two observations are worth emphasizing. First, anisotropic dissipation alone is already sufficient to give direction-dependent curve-shape transport-extracted temperatures from $r_{\mathrm{lin}}$--$T$ curves. This is seen most clearly at $q=0.5$, where the Josephson couplings are isotropic and the equilibrium model has no directional anisotropy. Nevertheless, Table~\ref{tab:anisR_RT} shows that $T_x^{\mathrm{HN}}(j)>T_y^{\mathrm{HN}}(j)$ for all three probe currents, and that $T_x^{1\%}>T_y^{1\%}$ at $j=0.001$. This indicates that the larger shunt resistance along the $x$ direction leads to a higher extracted temperature from the $r_{\mathrm{lin}}$--$T$ curves, even though the underlying thermodynamic transition is isotropic. This further supports the conclusion that a directional separation of transport-extracted temperatures does not by itself imply a split equilibrium BKT temperature or thermodynamic transition. 

Second, as shown in Fig.~\ref{fig:comparert}, anisotropic dissipation can further enhance the apparent splitting extracted from the low-resistance threshold criterion when the larger shunt resistance is assigned to the more strongly coupled direction. For the $1\%$ threshold at $j=0.001$, we find
\begin{equation}\label{eq:1percent_anisR}
 \left.\frac{T_y^{1\%}}{T_x^{1\%}}\right|_{q=0.6}=0.972,
 \qquad
 \left.\frac{T_y^{1\%}}{T_x^{1\%}}\right|_{q=0.7}=0.948.
\end{equation}
These ratios are smaller than those in Eq.~\eqref{eq:1percent_isoR}, meaning that the corresponding directional separation is enhanced by anisotropic dissipation. Equivalently, the relative splitting $1-T_y^{1\%}/T_x^{1\%}$ increases from $1.6\%$ to $2.8\%$ for $q=0.6$, and from $3.4\%$ to $5.2\%$ for $q=0.7$. On the other hand, the critical-scaling criteria continue to yield a single extracted temperature.

\section{Summary and discussion}\label{sec:discussion}
In this work, we have established a clean theoretical baseline for the transport phenomenology of anisotropic two-dimensional superconductors. By studying a minimal anisotropic JJA model in both equilibrium and nonequilibrium, we demonstrated that a system with a single thermodynamic BKT transition can nonetheless exhibit direction-dependent transport-extracted temperatures in operational transport metrics. This provides a minimal route to an apparent ``double-$T_c$'' without splitting the thermodynamic transition.

Within this minimal model, the apparent double-$T_c$ has a clear physical origin: anisotropic phase stiffness and anisotropic dissipation (vortex viscosity) naturally reshape the $r_{\mathrm{lin}}$--$T$ curves differently for different current directions. Because finite system sizes and finite measurement currents cut off the critical divergence of the correlation length, experimental and numerical fits are forced into a higher-temperature crossover regime. In this regime, standard curve-shape criteria—such as fixed-resistance thresholds or Halperin--Nelson fits—inherit this anisotropy, producing a spurious splitting of the curve-shape transport-extracted temperatures. 

Crucially, we found a stark contrast between curve-shape criteria and critical scaling criteria. Dynamic finite-size scaling and the $\alpha=3$ exponent in the simulated $E$--$j$ characteristics, equivalently the $I$--$V$ exponent in experimental notation, inherently account for finite-size cutoffs and are tied to the universal stiffness jump. Consequently, within the numerical resolution of the simulations, these criteria are much less sensitive to the crossover anisotropy and remain consistent with the single equilibrium $T_{\mathrm{BKT}}$.

This contrast serves as a useful diagnostic comparison for experiments. In our clean, minimal model, trivial anisotropy yields an apparent $R$--$T$ splitting of roughly $5\%$ (e.g., $T_y^{1\%} / T_x^{1\%} \approx 0.948$ for $q=0.7$ with anisotropic dissipation), while maintaining a unified $\alpha=3$ crossing. This comparison also clarifies the scope of our baseline model. Recent experiments on EuO/KTaO$_3$(111) interfaces \cite{Huang2026} report a much larger $T_c$ splitting ($\sim 20\%$) that persists even in the $\alpha=3$ criterion. Such behavior is not reproduced by the clean anisotropic phase-only model studied here. It therefore points to physics beyond this minimal baseline, although the present work alone does not determine the microscopic origin of that physics. Possible ingredients include, for example, fractional or half-vortices, directed pinning or disorder, intrinsic normal-state anisotropy, multiorbital effects, magnetic proximity effects, or stripe-like spatial textures \cite{hua2024superconducting,li2024theory,xu2025anisotropic,Huang2026}.

Ultimately, before invoking exotic physics, apparent splittings extracted only from $R$--$T$ curve-shape criteria should first be tested against anisotropic crossover effects. By systematically comparing curve-shape criteria from $R$--$T$ curves with critical-scaling criteria from nonlinear $I$--$V$ characteristics, experimentalists can cleanly distinguish trivial transport artifacts from genuinely new superconducting phases of matter. 

\begin{acknowledgments}
This work is supported by the National Key R\&D Program of China (Grant No. 2022YFA1403201) and the National Natural Science Foundation of China (Grant No. 12074411). We thank Ziji Xiang, Zi-Xiang Li, Kun Jiang, Hong Yao, Gaurav Khairnar and Zhipeng Xu for helpful discussions.
\end{acknowledgments}

\section*{Data Availability}
The data that support the findings of this article are available from the authors upon reasonable request. 

\appendix
\section{The Directional Helicity Modulus}\label{app:helicity_modulus}
\subsection{Derivation of Equation~\eqref{eq:helicity_modulus_MC}}\label{app:helicity_modulus_derivation}
Starting from Eq.~\eqref{eq:Fisher_formula}, we fix the direction of the extra twist to be $\nu=x,y$ and write the free energy as
\begin{equation*}
F = -T \ln Z, \qquad Z = \int \mathcal{D}\theta \, \exp[-\beta H(\theta, \Phi_\nu)],
\end{equation*}
where $H(\theta, \Phi_\nu)$ incorporates a uniform twist per bond $\delta_\nu \equiv \Phi_\nu/L$ only on bonds parallel to the chosen direction $\nu$: 
\begin{equation}
H(\theta, \Phi_\nu) = - \sum_i \sum_{\mu=x,y} E_{J,\mu} \cos\!\left(\theta_{i+\hat{\mu}}-\theta_i+\delta_\nu\delta_{\mu\nu}\right).
\end{equation}
The first derivative of the free energy with respect to $\delta_\nu$ is
\begin{equation*}
\frac{\partial F}{\partial \delta_\nu} = -T \frac{1}{Z} \frac{\partial Z}{\partial \delta_\nu} = \left\langle \frac{\partial H}{\partial \delta_\nu} \right\rangle,
\end{equation*}
where
\begin{equation}
\frac{\partial H}{\partial \delta_\nu} = \sum_i E_{J,\nu}\sin(\theta_{i+\hat{\nu}}-\theta_i+\delta_\nu).
\end{equation}
The second derivative is
\begin{equation*}
\frac{\partial^2 F}{\partial \delta_\nu^2}=\left\langle \frac{\partial^2 H}{\partial \delta_\nu^2} \right\rangle - \beta \left[ \left\langle \left( \frac{\partial H}{\partial \delta_\nu} \right)^2 \right\rangle - \left\langle \frac{\partial H}{\partial \delta_\nu} \right\rangle^2 \right].
\end{equation*}
The second derivative of the Hamiltonian is
\begin{equation}
\frac{\partial^2 H}{\partial \delta_\nu^2} = \sum_i E_{J,\nu}\cos(\theta_{i+\hat{\nu}}-\theta_i+\delta_\nu).
\end{equation}
Evaluating these expressions at $\delta_\nu = 0$ and then relabeling $\nu\to\mu$ gives Eq.~\eqref{eq:helicity_modulus_MC}:
\begin{equation*}
\Upsilon_\mu(L,T) = \frac{1}{L^2} \left. \frac{\partial^2 F}{\partial \delta_\mu^2} \right|_{\delta_\mu=0} = \frac{1}{L^2} \langle C_\mu \rangle - \frac{\beta}{L^2} \left[ \langle S_\mu^2 \rangle - \langle S_\mu \rangle^2 \right].
\end{equation*}

\subsection{Remark on finite boundary twists}\label{app:helicity_modulus_finite_twist}

The helicity modulus in the main text (Eq.~\eqref{eq:Fisher_formula}) is defined at zero boundary twist and is the quantity that enters the equilibrium BKT criterion, Eq.~\eqref{eq:Helicity_jump}. 

For a \emph{finite} boundary twist $\Theta_\mu = A_\mu L \neq 0$, one can define two similar objects: 
\begin{enumerate}
	\item the phase stiffness defined in Ref.~\cite{khairnar2025helicity} from the free-energy difference, 
		\begin{equation}
			\Upsilon_\mu^{(1)}(L,T;\Theta_\mu)
			\equiv \frac{2[F(\Theta_\mu)-F(0)]}{\Theta_\mu^{2}},
		\end{equation}
	\item the second derivative of the free energy at a nonzero boundary twist, 
		\begin{equation}\label{eq:upsilon2_finite_twist}
			\Upsilon_\mu^{(2)}(L,T;{\Theta}_\mu)
			\equiv \left.\frac{1}{L^2} \frac{\partial^2 F\left(\Theta_\mu+L \delta_\mu\right)}{\partial\delta_\mu^2}\right|_{\delta_\mu=0}, 
		\end{equation}
		which also admits an expression analogous to Eq.~\eqref{eq:helicity_modulus_MC} but with averages taken in the finite-twisted ensemble. 
\end{enumerate}

These need not coincide with the standard $\Upsilon_\mu$ defined under zero boundary twist. Indeed, Ref.~\cite{khairnar2025helicity} demonstrated that in the two-dimensional XY model $\Upsilon^{(1)}$ (denoted there as $\rho_s(\Theta)$) deviates from the helicity modulus defined under zero boundary twist when $\Theta_\mu$ is finite, owing to the mixing of states with opposite chirality. The same logic implies that $\Upsilon_\mu^{(2)}$ evaluated at a finite background twist should not be regarded as the standard helicity modulus. 

For the present problem, the reason finite twists cannot split the thermodynamic BKT transition is straightforward. When the total twist $\Theta_\mu$ is held fixed while $L\to\infty$, the corresponding gauge field scales as $A_\mu = \Theta_\mu/L \to 0$. At the spin-wave level we have 
\begin{equation}
 \frac{\Delta F_{\rm sw}}{L^2} \simeq \frac{1}{2}\Upsilon_\mu(\Theta_\mu/L)^2 = \mathcal O(L^{-2}). 
\end{equation}
Thus a finite boundary twist modifies only boundary and finite-size contributions, leaving the bulk long-wavelength theory unchanged. The renormalization-group flow therefore remains governed by a single geometric-mean stiffness.

\begin{figure}[tb]
	\centering
	\includegraphics[width=1.\linewidth]{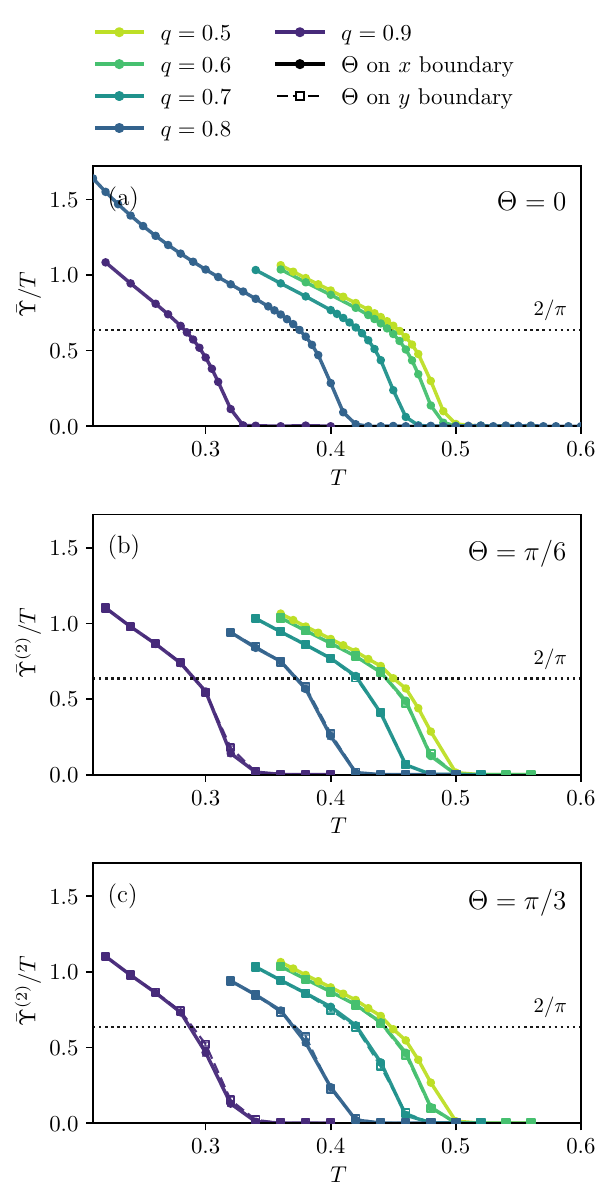}
	\caption{Geometric-mean stiffness response divided by temperature with system size $L=200$ and boundary twists: (a) $\Theta=0$, (b) $\Theta=\pi/6$, and (c) $\Theta=\pi/3$. For finite twist panels (b) and (c), we evaluate $\bar\Upsilon^{(2)}=\sqrt{\Upsilon_x^{(2)}\Upsilon_y^{(2)}}$ with $\Upsilon_\mu^{(2)}$ defined by Eq.~\eqref{eq:upsilon2_finite_twist}. The solid circles and open squares in (b) and (c) correspond to $\left(\Theta_x, \Theta_y\right)=\left(\Theta,0\right)$ and $\left(0,\Theta\right)$. The dashed line indicates $2/\pi$. }
	\label{fig:finite_twist}
\end{figure}

As a consistency check, we have evaluated $\bar\Upsilon^{(2)}$ at $\left(\Theta_x, \Theta_y\right)=\left(\pi/6,0\right),\left(0,\pi/6\right), \left(\pi/3,0\right), \left(0,\pi/3\right)$ for $q=0.5,0.6,0.7,0.8,0.9$ up to $L=200$ by Monte Carlo simulations. Ref.~\cite{khairnar2025helicity} has pointed out that under a finite twist, the standard Wolff cluster update is no longer appropriate and one may employ Metropolis sweeps\cite{metropolis1949monte,metropolis1953equation}. This stems from the fact that when the boundary twist is finite, the bond energy is not preserved under the cluster flip 
\[
	\theta_i \mapsto \theta_i^* =2\alpha - \theta_i, 
\]
where $\alpha$ is the random angle and $\theta_i$ is in the cluster. However, we can use a modified cluster update: 
\begin{equation}\label{eq:pi-shift}
	\theta_i \mapsto \theta_i^* = \theta_i + \pi. 
\end{equation}
Unlike the standard reflection update, this ``$\pi$-shift'' leaves the gauge-invariant phase difference of every internal cluster bond unchanged, even in the presence of a finite background twist. The remaining boundary bonds are treated by the usual Wolff-type acceptance construction, so that detailed balance is maintained. However, the shift changes phases only by $\pi$ and is not ergodic by itself. We consequently combine it with Metropolis sweeps, which restore ergodicity while retaining the acceleration from the cluster algorithm. 

As shown in Fig.~\ref{fig:finite_twist}, within numerical resolution, these finite-twist checks show only a single crossing with $2/\pi$ for each tested anisotropy. Moreover, in Figs.~\ref{fig:finite_twist}(b) and~\ref{fig:finite_twist}(c), applying the same total boundary twist along $x$ direction $\left(\Theta_x, \Theta_y\right)=\left(\Theta,0\right)$ and $y$ direction $\left(\Theta_x, \Theta_y\right)=\left(0,\Theta\right)$ results in nearly overlapping curves. This indicates that the finite-twist response does not show a robust direction-dependent splitting. Therefore, finite boundary twists do not generate a second thermodynamic transition in our anisotropic model. 

\section{The RSJD equations under FTBC}\label{app:RSJD_EoM}

The total current $I_{i,\mu}$ on the bond connecting site $i$ to $i+\hat\mu$ is given by Eq.~(\ref{eq:bond_current}). Current conservation at each site requires
\begin{equation}\label{eq:current_conservation}
\sum_{\mu=x,y}\left(I_{i,\mu}-I_{i-\hat\mu,\mu}\right)=0.
\end{equation}
Following the FTBC formulation \cite{lobb1983theoretical,chung1989dynamical}, the spatially averaged current must match the external applied current density $I_{\mathrm{ext},\mu}^{\mathrm{phys}}$, 
\begin{equation}\label{eq:current_average}
\frac{1}{L^2}\sum_i I_{i,\mu}=I_{\mathrm{ext},\mu}^{\mathrm{phys}},
\qquad \mu=x,y.
\end{equation}

To make the equations dimensionless, we introduce a reference energy scale $E_*$ and a reference resistance $R_*$, and define the scales for current, time, and temperature as:
\begin{equation}
I_*\equiv\frac{2e}{\hbar}E_*,
\qquad
\tau_*\equiv\frac{\hbar}{2eI_*R_*},
\qquad
T_*\equiv\frac{E_*}{k_B}.
\end{equation}
Dimensionless variables are then defined by
\begin{equation}
\tau\equiv\frac{t}{\tau_*},
\qquad
j_\mu\equiv\frac{I_{\mathrm{ext},\mu}^{\mathrm{phys}}}{I_*}.
\end{equation}
The dimensionless total bond current is $\mathcal J_{i,\mu}=I_{i,\mu}/I_*$. Using the dimensionless parameters $J_\mu$ and $r_{0,\mu}$ defined in the main text, $\mathcal J_{i,\mu}$ becomes
\begin{equation}
\mathcal J_{i,\mu}=J_\mu\sin\phi_{i,\mu}+\frac{1}{r_{0,\mu}}\dot\phi_{i,\mu}+\zeta_{i,\mu},
\end{equation}
where $\zeta_{i,\mu}=\eta_{i,\mu}/I_*$ is the dimensionless noise. The phase difference is $\phi_{i,\mu}=\theta_{i+\hat\mu}-\theta_i-\Delta_\mu$, so $\dot\phi_{i,\mu}=\dot\theta_{i+\hat\mu}-\dot\theta_i-\dot\Delta_\mu$.

Substituting $\mathcal J_{i,\mu}$ into the dimensionless form of Eq.~(\ref{eq:current_conservation}) gives
\begin{equation}\label{eq:dimensionless_current_conservation}
-\sum_{\mu=x,y}\frac{1}{r_{0,\mu}}
\left(\dot\theta_{i+\hat\mu}-2\dot\theta_i+\dot\theta_{i-\hat\mu}\right)
=S_i,
\end{equation}
where the source term is
\begin{equation}\label{eq:source_term}
S_i=\sum_{\mu=x,y}
\left[
\mathcal J_{i,\mu}^{\mathrm{sc}}-\mathcal J_{i-\hat\mu,\mu}^{\mathrm{sc}}
+\zeta_{i,\mu}-\zeta_{i-\hat\mu,\mu}
\right].
\end{equation}
The terms proportional to $\dot\Delta_\mu$ cancel out in the site current conservation because they are spatially uniform. Equation~(\ref{eq:dimensionless_current_conservation}) can be inverted using the weighted Laplacian operator
\[
D^{(r_0)}_{ij}=\sum_\mu\frac{1}{r_{0,\mu}}(\delta_{j,i+\hat\mu}-2\delta_{j,i}+\delta_{j,i-\hat\mu})
\]
and the Green's function of $-D^{(r_0)}$,
\[
\sum_k[-D^{(r_0)}]_{ik}G_{kj}^{(r_0)}=\delta_{ij}-\frac{1}{L^2}.
\]
With this convention, $-D^{(r_0)}\dot\theta=S$ gives $\dot\theta_i=\sum_jG_{ij}^{(r_0)}S_j$, which yields Eq.~(\ref{eq:RSJD_EoM_theta_dimensionless}). Since $\sum_i S_i=0$, the uniform mode $\sum_i\dot\theta_i$ is decoupled and we can set it to zero.

To obtain the equation for the twist variables $\Delta_\mu$, we substitute $\mathcal J_{i,\mu}$ into the dimensionless form of Eq.~(\ref{eq:current_average}):
\begin{equation*}
\frac{1}{L^2}\sum_i
\left[
\mathcal J_{i,\mu}^{\mathrm{sc}}
+\frac{1}{r_{0,\mu}}\left(\dot\theta_{i+\hat\mu}-\dot\theta_i-\dot\Delta_\mu\right)
+\zeta_{i,\mu}
\right]
=j_\mu.
\end{equation*}
Under periodic boundary conditions for $\theta_i$, the sum $\sum_i(\dot\theta_{i+\hat\mu}-\dot\theta_i)$ vanishes. Solving for $\dot\Delta_\mu$ then directly yields Eq.~(\ref{eq:RSJD_EoM_delta_dimensionless}).

\bibliographystyle{apsrev4-2}
\bibliography{2DAnisoTc}

@article{metropolis1953equation,
  author  = {Metropolis, Nicholas and Rosenbluth, Arianna W. and Rosenbluth, Marshall N. and Teller, Augusta H. and Teller, Edward},
  title   = {Equation of State Calculations by Fast Computing Machines},
  journal = {The Journal of Chemical Physics},
  year    = {1953},
  volume  = {21},
  number  = {6},
  pages   = {1087--1092},
  month   = jun,
  doi     = {10.1063/1.1699114}
}

@article{wolff1989collective,
  title = {Collective Monte Carlo Updating for Spin Systems},
  author = {Wolff, Ulli},
  journal = {Phys. Rev. Lett.},
  volume = {62},
  number = {4},
  pages = {361--364},
  year = {1989},
  month = {Jan},
  publisher = {American Physical Society},
  doi = {10.1103/PhysRevLett.62.361},
  url = {https://link.aps.org/doi/10.1103/PhysRevLett.62.361}
}

@article{hua2024superconducting,
  author  = {Hua, Xiangyu and Zeng, Zimeng and Meng, Fanbao and Yao, Hongxu and Huang, Zongyao and Long, Xuanyu and Li, Zhaohang and Wang, Youfang and Wang, Zhenyu and Wu, Tao and Weng, Zhengyu and Wang, Yihua and Liu, Zheng and Xiang, Ziji and Chen, Xianhui},
  title   = {Superconducting stripes induced by ferromagnetic proximity in an oxide heterostructure},
  journal = {Nature Physics},
  year    = {2024},
  volume  = {20},
  number  = {6},
  pages   = {957--963},
  month   = jun,
  issn    = {1745-2481},
  doi     = {10.1038/s41567-024-02443-x},
  url     = {https://doi.org/10.1038/s41567-024-02443-x}
}

@article{ambegaokar1980dynamics,
  title = {Dynamics of superfluid films},
  author = {Ambegaokar, Vinay and Halperin, B. I. and Nelson, David R. and Siggia, Eric D.},
  journal = {Phys. Rev. B},
  volume = {21},
  number = {5},
  pages = {1806--1826},
  year = {1980},
  month = {Mar},
  publisher = {American Physical Society},
  doi = {10.1103/PhysRevB.21.1806},
  url = {https://link.aps.org/doi/10.1103/PhysRevB.21.1806}
}

@article{ambegaokar1978dissipation,
  title = {Dissipation in Two-Dimensional Superfluids},
  author = {Ambegaokar, Vinay and Halperin, B. I. and Nelson, David R. and Siggia, Eric D.},
  journal = {Phys. Rev. Lett.},
  volume = {40},
  number = {12},
  pages = {783--786},
  year = {1978},
  month = {Mar},
  publisher = {American Physical Society},
  doi = {10.1103/PhysRevLett.40.783},
  url = {https://link.aps.org/doi/10.1103/PhysRevLett.40.783}
}

@misc{li2024theory,
      title={Theory of an infinitely anisotropic phase of a two-dimensional superconductor}, 
      author={Zi-Xiang Li and Steven A. Kivelson and Dung-Hai Lee},
      year={2024},
      eprint={2407.10269},
      archiveprefix={arXiv},
      primaryclass={cond-mat.supr-con},
      url={https://arxiv.org/abs/2407.10269}, 
}

@misc{xu2025anisotropic,
      title={Anisotropic vortex motion and two-dimensional superconducting transition}, 
      author={Zhipeng Xu and Kun Jiang and Jiangping Hu},
      year={2025},
      eprint={2506.05830},
      archiveprefix={arXiv},
      primaryclass={cond-mat.supr-con},
      url={https://arxiv.org/abs/2506.05830}, 
}

@article{kosterlitz1974critical,
doi = {10.1088/0022-3719/7/6/005},
url = {https://doi.org/10.1088/0022-3719/7/6/005},
year = {1974},
month = mar,
volume = {7},
number = {6},
pages = {1046},
author = {J M Kosterlitz},
title = {The critical properties of the two-dimensional xy model},
journal = {Journal of Physics C: Solid State Physics}
}

@article{khairnar2025helicity,
  title = {Helicity modulus and chiral symmetry breaking for boundary conditions with finite twist},
  author = {Khairnar, Gaurav and Vojta, Thomas},
  journal = {Phys. Rev. E},
  volume = {111},
  number = {2},
  pages = {024114},
  year = {2025},
  month = {Feb},
  publisher = {American Physical Society},
  doi = {10.1103/PhysRevE.111.024114},
  url = {https://link.aps.org/doi/10.1103/PhysRevE.111.024114}
}

@article{fisher1973helicity,
  title = {Helicity Modulus, Superfluidity, and Scaling in Isotropic Systems},
  author = {Fisher, Michael E. and Barber, Michael N. and Jasnow, David},
  journal = {Phys. Rev. A},
  volume = {8},
  number = {2},
  pages = {1111--1124},
  year = {1973},
  month = {Aug},
  publisher = {American Physical Society},
  doi = {10.1103/PhysRevA.8.1111},
  url = {https://link.aps.org/doi/10.1103/PhysRevA.8.1111}
}

@article{medvedyeva2000analysis,
  title = {Analysis of current-voltage characteristics of two-dimensional superconductors: Finite-size scaling behavior in the vicinity of the Kosterlitz-Thouless transition},
  author = {Medvedyeva, Kateryna and Kim, Beom Jun and Minnhagen, Petter},
  journal = {Phys. Rev. B},
  volume = {62},
  number = {21},
  pages = {14531--14540},
  year = {2000},
  month = {Dec},
  publisher = {American Physical Society},
  doi = {10.1103/PhysRevB.62.14531},
  url = {https://link.aps.org/doi/10.1103/PhysRevB.62.14531}
}

@article{medvedyeva2001ubiquitous,
title = {Ubiquitous finite-size scaling features in I–V characteristics of various dynamic XY models in two dimensions},
journal = {Physica C: Superconductivity},
volume = {355},
number = {1},
pages = {6--14},
year = {2001},
issn = {0921-4534},
doi = {10.1016/S0921-4534(01)00026-0},
url = {https://www.sciencedirect.com/science/article/pii/S0921453401000260},
author = {Kateryna Medvedyeva and Kim, Beom Jun and Petter Minnhagen}
}

@article{kim1999vortex,
  title = {Vortex dynamics for two-dimensional $\mathrm{XY}$ models},
  author = {Kim, Beom Jun and Minnhagen, Petter and Olsson, Peter},
  journal = {Phys. Rev. B},
  volume = {59},
  number = {17},
  pages = {11506--11522},
  year = {1999},
  month = {May},
  publisher = {American Physical Society},
  doi = {10.1103/PhysRevB.59.11506},
  url = {https://link.aps.org/doi/10.1103/PhysRevB.59.11506}
}

@article{olsson1995monte1,
  title = {Monte Carlo analysis of the two-dimensional XY model. I. Self-consistent boundary conditions},
  author = {Olsson, Peter},
  journal = {Phys. Rev. B},
  volume = {52},
  number = {6},
  pages = {4511--4525},
  year = {1995},
  month = {Aug},
  publisher = {American Physical Society},
  doi = {10.1103/PhysRevB.52.4511},
  url = {https://link.aps.org/doi/10.1103/PhysRevB.52.4511}
}

@article{olsson1995monte2,
  title = {Monte Carlo analysis of the two-dimensional XY model. II. Comparison with the Kosterlitz renormalization-group equations},
  author = {Olsson, Peter},
  journal = {Phys. Rev. B},
  volume = {52},
  number = {6},
  pages = {4526--4535},
  year = {1995},
  month = {Aug},
  publisher = {American Physical Society},
  doi = {10.1103/PhysRevB.52.4526},
  url = {https://link.aps.org/doi/10.1103/PhysRevB.52.4526}
}

@article{choi2000boundary,
  title = {Boundary effects on dynamic behavior of Josephson-junction arrays},
  author = {Choi, M. Y. and Jeon, Gun Sang and Yoon, Mina},
  journal = {Phys. Rev. B},
  volume = {62},
  number = {9},
  pages = {5357--5360},
  year = {2000},
  month = {Sep},
  publisher = {American Physical Society},
  doi = {10.1103/PhysRevB.62.5357},
  url = {https://link.aps.org/doi/10.1103/PhysRevB.62.5357}
}

@article{chung1989dynamical,
  title = {Dynamical properties of superconducting arrays},
  author = {Chung, J. S. and Lee, K. H. and Stroud, D.},
  journal = {Phys. Rev. B},
  volume = {40},
  number = {10},
  pages = {6570--6580},
  year = {1989},
  month = {Oct},
  publisher = {American Physical Society},
  doi = {10.1103/PhysRevB.40.6570},
  url = {https://link.aps.org/doi/10.1103/PhysRevB.40.6570}
}

@article{golubov2004current,
  title = {The current-phase relation in Josephson junctions},
  author = {Golubov, A. A. and Kupriyanov, M. Yu. and Il'ichev, E.},
  journal = {Rev. Mod. Phys.},
  volume = {76},
  number = {2},
  pages = {411--469},
  year = {2004},
  month = {Apr},
  publisher = {American Physical Society},
  doi = {10.1103/RevModPhys.76.411},
  url = {https://link.aps.org/doi/10.1103/RevModPhys.76.411}
}

@article{berezinskii1971destruction,
  author = {Berezinskii, V. L.},
  title = {Destruction of Long-range Order in One-dimensional and Two-dimensional Systems having a Continuous Symmetry Group I. Classical Systems},
  journal = {Sov. Phys. JETP},
  volume = {32},
  number = {3},
  pages = {493--500},
  year = {1971},
}

@article{kosterlitz1973ordering,
  author = {Kosterlitz, J. M. and Thouless, D. J.},
  title = {Ordering, metastability and phase transitions in two-dimensional systems},
  journal = {Journal of Physics C: Solid State Physics},
  volume = {6},
  number = {7},
  pages = {1181--1203},
  year = {1973},
  doi = {10.1088/0022-3719/6/7/010},
  url = {https://doi.org/10.1088/0022-3719/6/7/010}
}

@article{nelson1977universal,
  title = {Universal Jump in the Superfluid Density of Two-Dimensional Superfluids},
  author = {Nelson, David R. and Kosterlitz, J. M.},
  journal = {Phys. Rev. Lett.},
  volume = {39},
  number = {19},
  pages = {1201--1205},
  year = {1977},
  month = {Nov},
  publisher = {American Physical Society},
  doi = {10.1103/PhysRevLett.39.1201},
  url = {https://link.aps.org/doi/10.1103/PhysRevLett.39.1201}
}

@article{beasley1979possibility,
  title = {Possibility of Vortex-Antivortex Pair Dissociation in Two-Dimensional Superconductors},
  author = {Beasley, M. R. and Mooij, J. E. and Orlando, T. P.},
  journal = {Phys. Rev. Lett.},
  volume = {42},
  number = {17},
  pages = {1165--1168},
  year = {1979},
  month = {Apr},
  publisher = {American Physical Society},
  doi = {10.1103/PhysRevLett.42.1165},
  url = {https://link.aps.org/doi/10.1103/PhysRevLett.42.1165}
}

@article{halperin1979resistive,
  author = {Halperin, B. I. and Nelson, D. R.},
  title = {Resistive transition in superconducting films},
  journal = {J. Low Temp. Phys.},
  volume = {36},
  number = {5},
  pages = {599--616},
  year = {1979},
  month = {sep},
  doi = {10.1007/BF00116988},
  url = {https://doi.org/10.1007/BF00116988}
}

@article{minnhagen1987two,
  title = {The two-dimensional Coulomb gas, vortex unbinding, and superfluid-superconducting films},
  author = {Minnhagen, Petter},
  journal = {Rev. Mod. Phys.},
  volume = {59},
  number = {4},
  pages = {1001--1066},
  year = {1987},
  month = {Oct},
  publisher = {American Physical Society},
  doi = {10.1103/RevModPhys.59.1001},
  url = {https://link.aps.org/doi/10.1103/RevModPhys.59.1001}
}

@article{fisher1991thermal,
  title = {Thermal fluctuations, quenched disorder, phase transitions, and transport in type-II superconductors},
  author = {Fisher, Daniel S. and Fisher, Matthew P. A. and Huse, David A.},
  journal = {Phys. Rev. B},
  volume = {43},
  number = {1},
  pages = {130--159},
  year = {1991},
  month = {Jan},
  publisher = {American Physical Society},
  doi = {10.1103/PhysRevB.43.130},
  url = {https://link.aps.org/doi/10.1103/PhysRevB.43.130}
}

@article{hubscher2013stiffness,
  title = {Stiffness jump in the generalized $XY$ model on the square lattice},
  author = {H\"ubscher, David M. and Wessel, Stefan},
  journal = {Phys. Rev. E},
  volume = {87},
  number = {6},
  pages = {062112},
  year = {2013},
  month = {Jun},
  publisher = {American Physical Society},
  doi = {10.1103/PhysRevE.87.062112},
  url = {https://link.aps.org/doi/10.1103/PhysRevE.87.062112}
}

@article{weber1988monte,
  title = {Monte Carlo determination of the critical temperature for the two-dimensional XY model},
  author = {Weber, Hans and Minnhagen, Petter},
  journal = {Phys. Rev. B},
  volume = {37},
  number = {10},
  pages = {5986--5989},
  year = {1988},
  month = {Apr},
  publisher = {American Physical Society},
  doi = {10.1103/PhysRevB.37.5986},
  url = {https://link.aps.org/doi/10.1103/PhysRevB.37.5986}
}

@article{mon1989phase,
  title = {Phase Coherence and Nonequilibrium Behavior in Josephson Junction Arrays},
  author = {Mon, K. K. and Teitel, S.},
  journal = {Phys. Rev. Lett.},
  volume = {62},
  number = {6},
  pages = {673--676},
  year = {1989},
  month = {Feb},
  publisher = {American Physical Society},
  doi = {10.1103/PhysRevLett.62.673},
  url = {https://link.aps.org/doi/10.1103/PhysRevLett.62.673}
}

@article{teitel1983phase,
  title = {Phase transitions in frustrated two-dimensional $\mathrm{XY}$ models},
  author = {Teitel, S. and Jayaprakash, C.},
  journal = {Phys. Rev. B},
  volume = {27},
  number = {1},
  pages = {598--601},
  year = {1983},
  month = {Jan},
  publisher = {American Physical Society},
  doi = {10.1103/PhysRevB.27.598},
  url = {https://link.aps.org/doi/10.1103/PhysRevB.27.598}
}

@article{ohta1979xy,
  title = {$\mathrm{XY}$ model and the superfluid density in two dimensions},
  author = {Ohta, Takao and Jasnow, David},
  journal = {Phys. Rev. B},
  volume = {20},
  number = {1},
  pages = {139--146},
  year = {1979},
  month = {Jul},
  publisher = {American Physical Society},
  doi = {10.1103/PhysRevB.20.139},
  url = {https://link.aps.org/doi/10.1103/PhysRevB.20.139}
}

@article{jose1977renormalization,
  title = {Renormalization, vortices, and symmetry-breaking perturbations in the two-dimensional planar model},
  author = {Jos\'e, Jorge V. and Kadanoff, Leo P. and Kirkpatrick, Scott and Nelson, David R.},
  journal = {Phys. Rev. B},
  volume = {16},
  number = {3},
  pages = {1217--1241},
  year = {1977},
  month = {Aug},
  publisher = {American Physical Society},
  doi = {10.1103/PhysRevB.16.1217},
  url = {https://link.aps.org/doi/10.1103/PhysRevB.16.1217}
}

@article{schultka1994finite,
  title = {Finite-size scaling in two-dimensional superfluids},
  author = {Schultka, Norbert and Manousakis, Efstratios},
  journal = {Phys. Rev. B},
  volume = {49},
  number = {17},
  pages = {12071--12077},
  year = {1994},
  month = {May},
  publisher = {American Physical Society},
  doi = {10.1103/PhysRevB.49.12071},
  url = {https://link.aps.org/doi/10.1103/PhysRevB.49.12071}
}

@article{ambegaokar1969voltage,
  title = {Voltage Due to Thermal Noise in the dc Josephson Effect},
  author = {Ambegaokar, Vinay and Halperin, B. I.},
  journal = {Phys. Rev. Lett.},
  volume = {22},
  number = {25},
  pages = {1364--1366},
  year = {1969},
  month = {Jun},
  publisher = {American Physical Society},
  doi = {10.1103/PhysRevLett.22.1364},
  url = {https://link.aps.org/doi/10.1103/PhysRevLett.22.1364}
}

@article{lobb1983theoretical,
  title = {Theoretical interpretation of resistive transition data from arrays of superconducting weak links},
  author = {Lobb, C. J. and Abraham, David W. and Tinkham, M.},
  journal = {Phys. Rev. B},
  volume = {27},
  number = {1},
  pages = {150--157},
  year = {1983},
  month = {Jan},
  publisher = {American Physical Society},
  doi = {10.1103/PhysRevB.27.150},
  url = {https://link.aps.org/doi/10.1103/PhysRevB.27.150}
}

@article{stewart1968current,
    author = {Stewart, W. C.},
    title = {CURRENT‐VOLTAGE CHARACTERISTICS OF JOSEPHSON JUNCTIONS},
    journal = {Applied Physics Letters},
    volume = {12},
    number = {8},
    pages = {277--280},
    year = {1968},
    month = {04},
    issn = {0003-6951},
    doi = {10.1063/1.1651991},
    url = {https://doi.org/10.1063/1.1651991},
}

@article{mccumber1968effect,
    author = {McCumber, D. E.},
    title = {Effect of ac Impedance on dc Voltage‐Current Characteristics of Superconductor Weak‐Link Junctions},
    journal = {Journal of Applied Physics},
    volume = {39},
    number = {7},
    pages = {3113--3118},
    year = {1968},
    month = {06},
    issn = {0021-8979},
    doi = {10.1063/1.1656743},
    url = {https://doi.org/10.1063/1.1656743},
}

@article{greenside1981numerical,
author = {Greenside, H. S. and Helfand, E.},
title = {Numerical Integration of Stochastic Differential Equations—II},
journal = {Bell System Technical Journal},
volume = {60},
number = {8},
pages = {1927--1940},
doi = {10.1002/j.1538-7305.1981.tb00303.x},
url = {https://onlinelibrary.wiley.com/doi/abs/10.1002/j.1538-7305.1981.tb00303.x},
year = {1981}
}

@misc{Huang2026,
  title = {Directional-dependent {Berezinskii-Kosterlitz-Thouless} transition at {EuO/KTaO$_3$(111)} interfaces},
  author = {Huang, Zongyao and Wang, Zhengjie and Hua, Xiangyu and Wang, Huiyu and Li, Zhaohang and Liu, Shihao and Wang, Zhiwei and Quan, Feixiong and Wang, Zhen and Tao, Jing and He, James Jun and Xiang, Ziji and Chen, Xianhui},
  year = {2026},
  month = {apr},
  eprint = {2604.00608},
  archiveprefix = {arXiv},
  primaryclass = {cond-mat.supr-con},
  url = {https://arxiv.org/abs/2604.00608}
}

@article{metropolis1949monte,
author = {Nicholas Metropolis and S. Ulam},
title = {The Monte Carlo Method},
journal = {Journal of the American Statistical Association},
volume = {44},
number = {247},
pages = {335--341},
year = {1949},
publisher = {Taylor \& Francis},
doi = {10.1080/01621459.1949.10483310},
note = {PMID: 18139350},
url = {https://www.tandfonline.com/doi/abs/10.1080/01621459.1949.10483310},
}

@article{helfand1979numerical,
author = {Helfand, E.},
title = {Numerical Integration of Stochastic Differential Equations},
journal = {Bell System Technical Journal},
volume = {58},
number = {10},
pages = {2289--2299},
doi = {10.1002/j.1538-7305.1979.tb02967.x},
url = {https://onlinelibrary.wiley.com/doi/abs/10.1002/j.1538-7305.1979.tb02967.x},
year = {1979}
}

@article{simkin1997finite,
  title = {Finite size and current effects on IV characteristics of Josephson junction arrays},
  author = {Simkin, M. V. and Kosterlitz, J. M.},
  journal = {Phys. Rev. B},
  volume = {55},
  number = {17},
  pages = {11646--11650},
  year = {1997},
  month = {May},
  publisher = {American Physical Society},
  doi = {10.1103/PhysRevB.55.11646},
  url = {https://link.aps.org/doi/10.1103/PhysRevB.55.11646}
}

@article{benfatto2009broadening,
  title = {Broadening of the Berezinskii-Kosterlitz-Thouless superconducting transition by inhomogeneity and finite-size effects},
  author = {Benfatto, L. and Castellani, C. and Giamarchi, T.},
  journal = {Phys. Rev. B},
  volume = {80},
  number = {21},
  pages = {214506},
  year = {2009},
  month = {Dec},
  publisher = {American Physical Society},
  doi = {10.1103/PhysRevB.80.214506},
  url = {https://link.aps.org/doi/10.1103/PhysRevB.80.214506}
}

@article{hohenberg1977theory,
  title = {Theory of dynamic critical phenomena},
  author = {Hohenberg, P. C. and Halperin, B. I.},
  journal = {Rev. Mod. Phys.},
  volume = {49},
  number = {3},
  pages = {435--479},
  year = {1977},
  month = {Jul},
  publisher = {American Physical Society},
  doi = {10.1103/RevModPhys.49.435},
  url = {https://link.aps.org/doi/10.1103/RevModPhys.49.435}
}

\end{document}